\documentstyle[aps,epsfig,multicol,times,color]{revtex}

\newcommand{\Endrule}{\vskip 3pt\noindent\hrule width 8.6cm\vskip 3pt}
\newcommand{\Beginrule}{\vskip 3pt\noindent\hbox{%
\vbox{\hbox to 9cm{\hfill}}\vbox{\hrule width 9cm}} \vskip 3pt}

\newcommand{\br}{{\bf r}}

\newcommand{\bK}{{\bf K}}
\newcommand{\bP}{{\bf P}}
\newcommand{\bn}{{\bf n}}

\newcommand{\StartTwoColumn}{\narrowtext}
\newcommand{\EndTwoColumn}{\widetext}

\newcommand{\TwoColumnPaper}{\renewcommand{\StartTwoColumn}{\begin{multicols}{2}}
\renewcommand{\EndTwoColumn}{\end{multicols}}
\renewcommand{\narrowtext}{\Beginrule\begin{multicols}{2}}
\renewcommand{\widetext}{\end{multicols}\Endrule}}

\newcommand{\Label}[1]{\label{#1}}
\newcommand{\rw}[1]{}
\def\DRAFT{%
\renewcommand{\Label}[1]{\label{##1}
{\hbox to 0cm{\textcolor{green}{\hss\em ##1\quad}}}}
\renewcommand{\rw}[1]{\vskip 10pt%
\noindent{\framebox{\textcolor{red}{New Material Needed}}}%
\par\noindent{\textcolor{red}{\em ##1}}\vskip 10pt}
\def\Input##1{\include{##1}}}

\def\include#1{\input{#1}}
\def\Item[#1]{\vskip 3pt\noindent #1 }
\def\ER{E_\Delta }
\def\Emax{E_R }
\def\BEC{Bose-Einstein condensate}
\def\GP{Gross-Pitaevskii}

\TwoColumnPaper
\begin{document}

\title{Quantum Kinetic Theory VI: The Growth of a Bose-Einstein 
Condensate}
\author{M.D.~Lee  and C.W.~Gardiner} 

\address{School of Chemical 
and Physical Sciences, Victoria University, Wellington, New Zealand}

\maketitle
\begin{abstract} 
A detailed analysis of the growth of a \BEC\ is given, based on 
quantum kinetic theory, in which we take account of the evolution of 
the occupations of lower trap levels, and of the full Bose-Einstein 
formula for the occupations of higher trap levels, as well as the Bose 
stimulated direct transfer of atoms to the condensate level introduced 
by Gardiner {\em et al.} We find good agreement with experiment at 
higher temperatures, but at lower temperatures the experimentally 
observed growth rate is somewhat more rapid.  We also confirm the 
picture of the ``kinetic'' region of evolution, introduced by Kagan 
{\em et al.}, for the time up to the initiation of the condensate.  
The behavior after initiation essentially follows our original growth 
equation, but with a substantially increased rate coefficient.

Our modelling of growth implicitly gives a model of the spatial shape 
of the condensate vapor system as the condensate grows, and thus 
provides an alternative to the present phenomenological fitting 
procedure, based on the sum of a zero-chemical potential vapor and a 
Thomas-Fermi shaped condensate.  Our method may give substantially 
different results for condensate numbers and temperatures obtained 
from phenomentological fits, and indicates the need for more 
systematic investigation of the growth dynamics of the condensate from 
a supersaturated vapor.
\end{abstract}

\pacs{PACS Nos. }

\StartTwoColumn

\section{Introduction}

Although the race to produce a \BEC\ was preceded by intense debate 
concerning the likely rate of its formation, the discovery that a 
\BEC\ of alkali atoms could be produced relatively simply 
\cite{JILAexpt,MITexpt,RICEexpt}, and that the growth time was of the 
order of one second, moved most theoretical activity into the 
investigation of the properties of the condensates so produced.  Since 
the production of the first \BEC\ there have been few theoretical 
investigations into condensate growth, and only one experiment 
\cite{MITgrowth} has made any measurements of growth rates.  Only the 
work of the present authors and co-workers, based on quantum kinetic 
theory, has made quantitative predictions on the growth rate of a \BEC 
.  This work started when we showed how to introduce the concept of 
stimulated condensate growth resulting from kinetic processes 
\cite{BosGro}, leading to a very simple formula for the growth rate.  
The MIT experiment \cite{MITgrowth} took the form of a verification of 
the validity of our theoretical prediction.  At the same time in 
\cite{NewestBosGro} we refined the basic concept of Bosonic 
stimulation to generate a less idealized theoretical picture, and to 
compare it with experiment.  These initial papers were of necessity 
brief, and developed neither the full theoretical justification on the 
numerical modeling nor the full range of possible comparison with the 
available experimental data.  In particular, no account was taken of 
the information available on the spatial distribution of the atoms in 
the vapor-condensate system as the condensate grows from the vapor.

This paper will therefore give the detailed justifications and a full 
range of comparison with experimental data.  Most particularly, we 
want to present a theoretically justifiable method of describing the 
condensate vapor system as it grows.  The absence of such a 
description has led to a phenomenological fitting of vapor profiles to 
a {\em zero chemical potential} \cite{MITgrowth,Hydrogen1} 
Bose-Einstein distributions, which may be an imperfect 
model whose results could well be misleading.

The theoretical description of condensate growth that we present is 
largely able to be viewed as a modification of the quantum Boltzmann 
equation, in which, however, explicit note is taken of the 
modification of the excitation spectrum by the existence of the 
condensate, including of course the fact that the lowest single 
particle excitation energy is the chemical potential $\mu_C(n_0)$ of 
the condensate of $n_0$ atoms.  Equilibrium arises as a result of the 
equality of the chemical potentials of uncondensed vapor and 
condensate, a picture which is rather similar to that normally adopted 
for chemical reactions.  The quantum Boltzmann equation  itself 
automatically provides the Bose stimulation, which makes transition 
rates into the condensate and other highly occupied levels achieve a 
speed which permits the production of the condensate in a finite 
time.  Without Bose stimulation, the production of a condensate of 
about 1,000,000 sodium atoms would take 30 hours, rather than the 
100ms observed.

At first glance it might appear that a description which appears to be 
based on the quantum Boltzmann equation would have nothing to say 
about condensate coherence or the origin of that coherence.  This is 
emphatically not the case---the kinetics of the transfer of the 
between energy levels {\em in a trap} requires the existence of a 
wavefunction for each energy level.  The condensate level has its own 
wavefunction, and this obeys the \GP\ equation.  The coherence arises 
because this level becomes macroscopically occupied.  There is no {\em 
precise} moment when one can say that the condensate initiates.  This 
picture applies in a trap, in which the energy levels about which we 
have been speaking are rather well separated.  The picture of a \BEC\ 
as developed in the middle part of this century as a part of condensed 
matter theory is of a homogeneous and thus infinitely extended 
system---a system for which the thermodynamic limit is achieved.  
Looked at from our viewpoint, this would be achieved by making the 
trap broader and ultimately flat.  There is a transition point where 
the trap becomes so flat that there is an occupation of the lowest 
quasiparticle levels which becomes comparable with the occupation of 
the condensate itself.  At this stage the traditional condensed matter 
picture becomes relevant, but this is not achieved in any traps 
presently in use.

\section{Model for Growth of a Condensate}


In this section, the formalism of Quantum Kinetic theory 
\cite{QK}, will be used to form a model of the growth of 
a trapped Bose-Einstein condensate.  
The Bose atoms are described by a second-quantized 
field, in the pseudopotential approximation; that is, we write
\begin{eqnarray}\Label{In1}
H=H_{\rm kin}+H_I + H_T,
\end{eqnarray}
where
\begin{eqnarray}\Label{In2}
H_{\rm kin}&=&\int d^3{\bf x}\,\psi ^{\dagger }({\bf x})
\left(- {\hbar ^2\over2m}\nabla ^2\right) \psi ({\bf x})   ,
\\ \Label{In3}
H_I&=&{u\over 2}\int d^3{\bf x}\psi ^{\dagger }({\bf x})
\psi ^{\dagger }( {\bf x})  
\psi ( {\bf x}) \psi ( {\bf x}) .  
\end{eqnarray}
and the term $ H_T$ arises from a trapping potential as 
\begin{eqnarray}\Label{In4}
H_T =  \int d^3 {\bf x}\,V_T({\bf x)}\psi^\dagger({\bf x})\psi({\bf x}).
\end{eqnarray}
The pseudopotential method is used---its validity for this kind of 
system has been justified in QKV---where $ u = 4\pi a\hbar^2/m$, and $ 
a$ is the $s$-wave scattering length arising from the interatomic 
potential.

The situation being considered is 
that of a vapor cloud confined in a trap in which the lower energy 
levels are not significantly populated, whilst the higher energy 
levels contain thermalized equilibrium populations, characterized by a 
temperature $T$ and chemical potential $\mu$, unstable against 
condensate formation.

This situation is likely to arise, to a degree of approximation, if a 
system, which is initially in equilibrium at a temperature slightly 
greater than the critical temperature, is cooled very suddenly to a 
temperature below the critical temperature, by removing the very high 
energy atoms in a rapid evaporative cooling `cut'.  The higher energy 
levels will very quickly come to their equilibrium distributions, 
since the difference between the distributions before and after the 
cut are quite small at the higher energies.  The lower levels will 
however be far from equilibrium and evolve to form a condensate.  This 
is possibly one of the easiest scenarios to model, it is also the 
situation investigated by the only detailed experimental 
study of the growth of a condensate  in a gas of 
$^{23}{\rm Na}$ atoms~\cite{MITgrowth}.

\subsection{The System}
The system is contained in a three-dimensional harmonic potential, 
characterized by the frequencies $\omega_x$, $\omega_y$ and 
$\omega_z$.  It will be useful to define the geometrical mean 
frequency of the trap as $\omega = (\omega_x\omega_y\omega_z)^{1/3}$.

\subsubsection{Effective potentials arising from mean-field effects}
In this system the energies and wavefunctions of the lower trap levels 
are quite strongly affected by the presence of the condensate, and the 
effect will of course change as the condensate grows.  In QKV it was 
shown that it is reasonable to account for this by introducing 
mean-field effects, which make the effective potentials depend on the 
occupations of the bands.  The situation is illustrated in 
Fig.\ref{fig-mean-field}.  As the condensate grows, it expels the 
vapor from the center of the trap, and this expulsion serves to reduce 
the mean-field of the vapor as experienced by the condensate.  The 
growth will be assumed to be so slow that the condensate- and 
non-condensate-bands are always in {\em thermal} equilibrium---that 
is, they will have a well defined temperature shared by both of them, 
but will not have the same chemical potential.  Growth therefore 
occurs as atoms are transferred from the vapor to the condensate, 
leading eventually to a unique chemical potential for the whole 
system.

\begin{figure}[t]
\epsfig{file=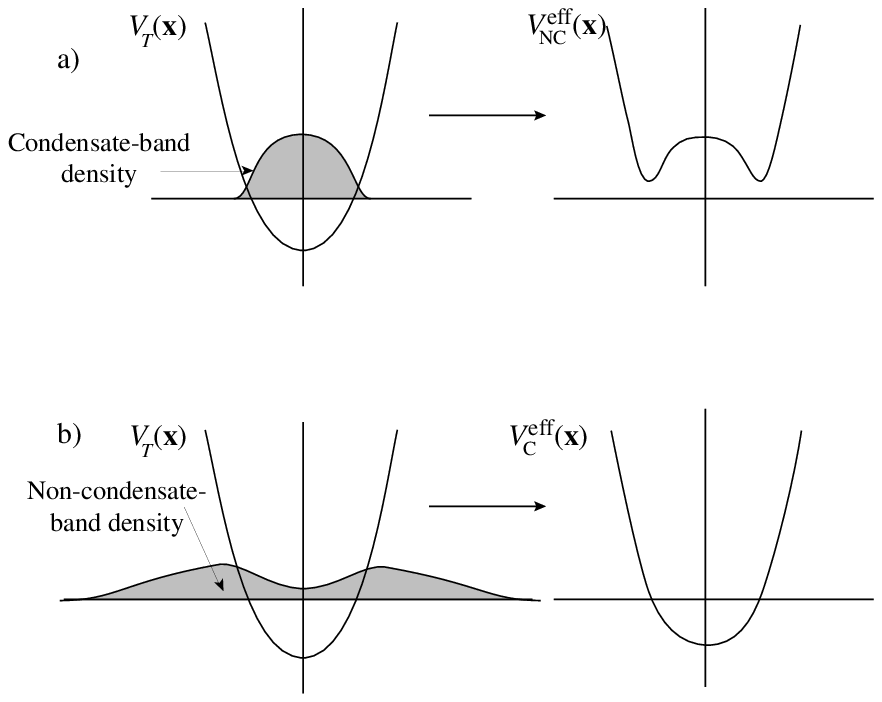,width=8.6cm}
\caption{Representation of the 
modification of the trapping potential for a) the non-condensate band 
and b) the condensate band due to mean-field effects.}
\Label{fig-mean-field}
\end{figure}

\subsubsection{Condensate- and non-condensate bands}
In the formalism of quantum kinetic theory, the system is divided into 
condensate and non-condensate bands.  We will treat in this paper the situation 
in which the non-condensate band is assumed to be in thermal equilibrium with a 
temperature $T$ and chemical potential $\mu$, and to contain the vast 
majority of the atoms so that it is essentially undepleted by the 
process of condensate formation.

The picture of growth  we will use is that presented in QKV.  In 
that paper, it was shown that a legitimate division into condensate 
and non-condensate bands can be made in which one distinguishes 
between {\em particle-like} excitations, to which it is possible to 
assign a definite number of atoms, and {\em phonon-like} excitations, 
which are collective modes, which normally involve a large average 
number of atoms, but are not eigenstates of the atom number. In 
practice, it has been shown \cite{StringariRev} that the energy above 
which all excitations are essentially particle-like is relatively 
small. 

For the purposes of our modeling, however, there are two criteria 
which must be considered in the definition of the condensate band.

\Item[i)] The non-condensate band is considered to be time 
independent, therefore the condensate band must include all levels 
whose populations change significantly during the condensate growth 
process.  For the non-condensate band, the thermal distribution is 
given in the bulk by $[e^{(E-\mu)/kT}-1]^{-1}$.  This is only valid 
for $E>\mu$, and gives very large populations when $E\approx \mu$.  
The transition rates in and out of levels in this vicinity also become 
very large, which contradicts the assumption that the distribution of 
the non-condensate band is time-independent.  These  lower states 
\emph{are} therefore must be treated time dependently, and hence must 
be included in the condensate band. 

\Item[ii)] The condensate band consists principally of levels whose 
energy eigenvalues are significantly affected by the presence of a 
condensate---but levels which are not affected may be included if this 
is desirable, which must be done if the first criterion is to be met.

Consequently, in this paper we will choose the condensate band to 
consist of all levels with energy less than the value $\Emax$.  We 
will also introduce an energy $\ER < \Emax$, which is the energy above 
which we can consider the energy levels to be unaffected by the 
condensate, as illustrated in Fig.\ref{fig:levelschange}.

\subsubsection{Grouping of energy levels into bands}
The inclusion of all the condensate band energy levels in the model 
means that simulations of the system require, in principle, the 
calculation of all the eigenfunctions of the condensate band, and 
detailed summations over these.  In practice the number of energy 
levels involved is of the order of tens of thousands, which makes an 
exact description impractical.  However, progress can be made by 
grouping together energy levels in the condensate band into small 
`sub-bands', with only the ground state (condensate state) being 
described as a single level.  Each sub-band is described by an energy 
$e_m$ and contains all the eigenstates found within the energy range 
$[e_m - \Delta e_m/2 , e_m +\Delta e_m/2]$.  The value of $\Delta e_m$ 
is chosen partially by the requirement that the lowest of these 
sub-bands contains at least 3 levels.  Smaller values of $\Delta e_m$ 
would lead sub-bands containing only fractions of individual levels, 
which is obviously unphysical.  As the condensate grows, the 
mean-field effects from the high occupation of the condensate level 
will cause the energies of the levels in the sub-bands to increase.  
The values of $e_m$ and $\Delta e_m$ are therefore dependent on the 
condensate occupation, and the manner in which they are altered will 
be discussed later.

\subsection{Notation}
For clarity, we set out some of our notation.

\begin{eqnarray}\Label{notation1}
&& N:  \mbox{number of atoms in the {\em condensate band}},
\\ \Label{notation2}
&&n_0:  \mbox{number of atoms in the {\em condensate}},
\\ \Label{notation5}
&&n_{0,f}:  \mbox{{\em equilibrium} number of atoms in the {\em 
condensate}},
\\ \Label{notation3}
&&\mu_C(n_0):  \mbox{chemical potential of the {\em condensate}},
\\ \Label{notation4}
&&\mu:  \mbox{chemical potential of the {\em non-condensate band}},
\\ \Label{notation6}
&&\xi_{n_0}({\bf x}): \mbox{wavefunction of an $n_0$ atom condensate}.
\end{eqnarray}
In the situations we will consider, the number of condensate atoms $ n_0$ will 
vary from zero to almost $ N$, but this will always be substantially less than 
the number of atoms in the whole system, composed of both condensate and 
non-condensate bands.  Thus, when the condensate is fully grown, the 
approximation $ n_0\approx N$ will be valid, and will often be used.

\subsection{Density of States for the System}
\Label{sec:densstates}

In the absence of any condensate, the density of states $G(E)$ is taken to be
that of a non-interacting gas in a harmonic well.  That is
\begin{equation}
G(E) \equiv {d{\rm N}(E) \over dE} = {(E-{3\over 2} \hbar \bar{\omega})^2
\over 2\hbar^3\omega_x\omega_y\omega_z}, \Label{densnonint}
\end{equation}
where ${\rm N}(E)$ is the cumulative number of states with energy less than $E$ 
and
$\bar{\omega} = (\omega_x+\omega_y+\omega_z)/3$. The number of states in the
sub-band with average energy $e_m$ is thus $g_m = G(e_m)\Delta e_m$.  The
energy scale is such that the value of $V_T(\br) = m(\omega_x^2x^2
+\omega_y^2y^2 +\omega_z^2z^2)/2$ is zero at the origin.

Once the condensate begins to form, the mean field effects need to be taken
into account.  The mean field repulsion due to the condensate changes the
energies of the lower trapped states.  The energy of the condensate level is
equal to the chemical potential $\mu_C(n_0)$, which increases with $n_0$.  In 
the
Thomas-Fermi approximation the rise is proportional to $n_0^{2/5}$.  Obviously
the condensate level must remain the lowest energy state, and thus the energies
of the other states below $\ER$ must also rise in some fashion.  The
exact nature of the energy change is difficult to calculate, but some
reasonable approximations can be proposed which should reproduce the
significant behavior caused by the mean field effects.  

The most simply calculated estimate of the energy changes is to assume the
energies of the sub-bands $e_m$ are evenly distributed between the fixed upper
limit of $\ER$ and the lower limit of $\mu_C(n_0) \approx \alpha 
n_0^{2/5}$. 
This is illustrated schematically in Fig.\ref{fig:levelschange}.  Both the
values of $e_m$ and the values of $\Delta e_m$ are now $n_0$ dependent. 

The final value of $\Delta e_m$ (after the growth of the condensate) 
is set to be $\hbar\omega$. This condition also always fulfills the 
requirements that there are at least about 3 discrete levels contained 
in the lowest energy sub-band, and yet ensures that the sub-band has 
only a relatively small energy range.

The density of states for the condensate band, in the presence of the
condensate, is thus taken to be approximately $G_{n_0}(E) = {\cal N}[E-
\mu_C({n_0})]^2$,
where ${\cal N}$ is a normalization chosen so that the cumulative number of
states at $\ER$ is the same as for the non-interacting harmonic
oscillator potential.  This behavior is illustrated in
Fig.\ref{fig:density}.  As the discontinuity at $\Emax$ shows, this
model is obviously quite simplistic, and more realistic models will be
discussed later.   

It should be noted that in the inset of Fig.\ref{fig:density} the number of
particles per energy interval $f(E)$ is shown (the occupation of the condensate
level is not shown) for equilibrium conditions.  From this inset it can be seen
that the vast majority of atoms do indeed reside at energies higher than
$\Emax$, and so the assumption that the non-condensate band is undepleted
should be valid.

\begin{figure}[t]
\begin{center}
\epsfig{file=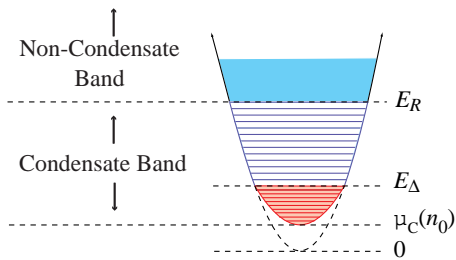,width=8.6cm}
\end{center}
\caption{Schematic representation of the simple model to change 
the energy levels due to the mean field effects of the condensate.  
Levels are evenly distributed between $\ER$ and $\mu_C({n_0})$ 
which increases with ${n_0}$.}
\Label{fig:levelschange}
\end{figure}

\subsubsection{Modified Thomas-Fermi Chemical Potential}
 In the Thomas Fermi approximation the chemical potential of the condensate is
given by
\begin{equation}
\mu_C({n_0}) = 
\left
({15a\omega_x\omega_y\omega_zm^{1/2}\hbar^2 \over 2^{5/2}}{n_0}\right)^{2/5},
\Label{TFmu2}
\end{equation}
which vanishes as ${n_0} \rightarrow 0$.  However, the ground state of a
non-interacting gas in a harmonic well is 
$\hbar(\omega_x+\omega_y+\omega_z)/2$, and so the real chemical potential
should approach this value as ${n_0} \rightarrow 0$.  In order to interpolate 
the Thomas-Fermi chemical potential to satisfy this requirement, the following 
form for the chemical potential will be used
\begin{equation}
\mu_C({n_0}) = \alpha({n_0} + \nu)^{2/5}
\Label{muN}
\end{equation}
where $\alpha =(15a\omega_x\omega_y\omega_z m^{1/2}\hbar^2/4\sqrt{2})^{2/5}$,
and $\nu$ is a constant such that $\alpha\nu^{2/5} =
\hbar(\omega_x+\omega_y+\omega_z)/2$.  

\subsubsection{Estimate of $\ER$}

An estimate for the value of $\ER$, above which the excitation 
spectrum is well described by that for a non-interacting gas, can be 
obtained using the number-conserving Bogoliubov 
spectrum~\cite{trueBog,QK3}.  The quantity of interest now is the ratio 
of the corrections to the energy level arising from the presence of 
the condensate, to the energy level determined by the non-interacting 
gas model.  For the case of the trap used in the $^{23}{\rm Na}$ 
growth experiments at MIT~\cite{MITgrowth}, numerical calculations 
show that the ratio is less than 10\% for 
energies  $ \ge 2\mu_C({n_0})$, and the 
corrections are of the order of only 5\% for $E
\sim 2.5\mu_C({n_0})$.  Thus a reasonable estimate of $\ER$ to be 
used in the simulations is
\begin{equation}
\ER = 2\mu_C(n_{0,f})
\end{equation}
where $n_{0,f}$ is the equilibrium occupation of the condensate level, 
and this is the value of $\ER$ which will be used in this 
paper. \Label{sec:Emax}

\begin{figure}[t]
\begin{center}
\epsfig{file=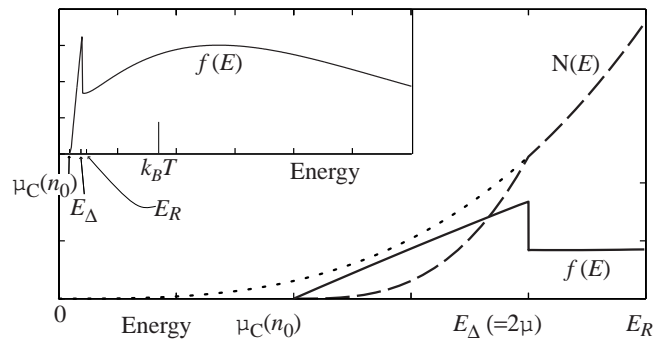,width=8.6cm}
\end{center}
\caption{The cumulative number of states for a gas in a harmonic potential well. 
The dotted line shows the situation for a non-interacting gas (no mean field
effects).  The dashed line represents the cumulative number of states due to
the simple model proposed in the text to incorporate mean field effects.  The
solid line shows the corresponding occupation per energy interval $f(E) =
G(E)[\exp((E-\mu)/kT) -1]^{-1}$, also shown on a larger scale in the inset.}
\Label{fig:density}
\end{figure}

\subsubsection{Comparison with More Accurate Density of States}
As a result of the predominantly single particle nature of the excitation
spectrum, the cumulative number of states ${\rm N}(E)$ is expected to be quite 
well
described at high energies by the semiclassical approximation
\begin{equation}
{\rm N}(E) = {1 \over {(2\pi\hbar)^3}}\int_0^E d\varepsilon
\; \int d{\bf r}\int d{\bf p} \ \delta (\varepsilon
-E_{\rm sp}({\bf p},{\bf r})) \;,
\Label{eq:semiclnstates}
\end{equation}
where 
$E_{\rm sp}({\bf p},{\bf r})=p^2/2m  +V_{T}({\bf r})
+(8\pi\hbar^2a/m)|\xi_{n_0}^2({\bf r})|-\mu_C({n_0})$ 
is the semiclassical energy of a
single particle in the potential created by the combination of the trapping
potential and the mean field repulsion of the condensate.
Equation~(\ref{eq:semiclnstates}) thus represents the summation over all phase
space cells which contain a single particle excitation state of energy less
than $E$. In~\cite{DGGPS1997}, by carrying out the momentum and space
integrals, an expression for equation~(\ref{eq:semiclnstates}) was found,
for the case of an isotropic harmonic trap of frequency $\omega$, in the form 
\begin{eqnarray}
\Label{stringarispec}
&&{{\rm N}(\tilde{E}) \over n_0} = 
 \tilde{E}^2 
\sqrt{({1 - x})(x + \eta/\tilde{E})}
\nonumber \\
&&\quad +
\int_0^{\tilde{E}} d\tilde{\varepsilon} 
{4 \over \pi \zeta(3)}
\int_0^1 dx\,  \tilde{\varepsilon} \eta
{\sqrt{{[x^2 +\tilde{\varepsilon}^2/\eta^2]^{1/2} - x} \over
{x^2 + \tilde{\varepsilon}^2/\eta^2}}} 
\nonumber\\
\end{eqnarray}
where
\begin{eqnarray}
\eta &=& {\mu_C(n_0) \over kT_c},\qquad
kT_c = \hbar\omega \left( {n_0 \over \zeta(3)}\right)^{1/3}.
\end{eqnarray}
The chemical potential $\mu_C(n_0) $ is given by the Thomas-Fermi approximate 
form (\ref{TFmu2}) and the
energy is given in the dimensionless units $\tilde{E} = E/kT_c$.  This
semiclassical form for ${\rm N}(E)$ was found in~\cite{DGGPS1997,StringariRev} 
to be
practically indistinguishable from that found by numerical solutions of the
Bogoliubov spectrum over the entire range of energies.

\begin{figure}[t]
 \epsfig{file=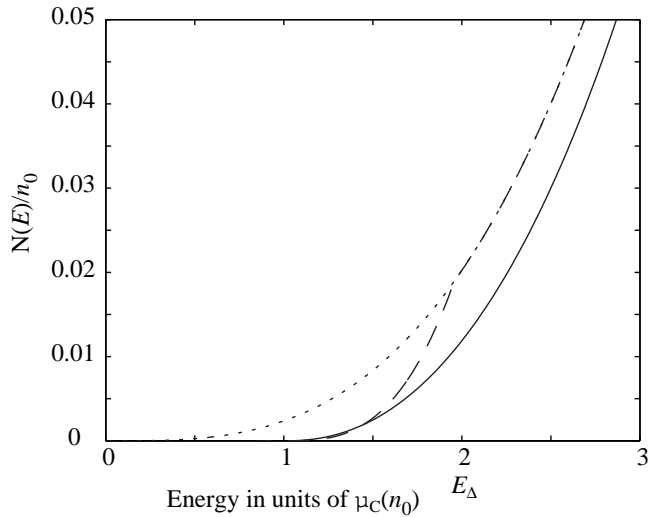,width=8.6cm} 

 \caption{The cumulative number of states ${\rm N}(E)$ below an energy 
 $E$.  The solid line shows the results obtained using the 
 semiclassical approximation for an isotropic trap 
 equation~(\ref{stringarispec}), the dotted line shows the case for 
 the non-interacting harmonic well.  The dashed line represents the 
 form used in this paper with $\ER = 2\mu_C(n_0)$.  The results were 
 obtained for a condensate of 5 million atoms at a temperature of 
 $900{\rm nK}$.}
 \Label{fig:stringarispec}
\end{figure}

In Fig.\ref{fig:stringarispec} the semiclassical form of ${\rm N}(E)$ 
obtained from equation~(\ref{stringarispec}) is compared to that of 
the non-interacting harmonic oscillator (given by 
equation~(\ref{densnonint})), and the density of states obtained using 
our model with $\ER$ equal to $2\mu_C(n_0)$.  The figure does not 
show very good agreement of our model with the semiclassical results 
at moderate energies, although at low energies the agreement is good.  
At high enough energies (not shown) the non-interacting potential 
results become practically indistinguishable from those obtained from 
the semiclassical method, the energy at which this occurs is about 
$5\mu_C(n_0)$ for the results in Fig.\ref{fig:stringarispec}.  It should be 
emphasized that this semiclassical result applies to an isotropic 
trap, and the consideration of the anisotropy of realistic traps, 
which has not yet been accounted for, may have a significant effect on 
the spectrum.

\subsubsection{Approximations}
Several approximations 
that have been made in the derivation of this model rely on the 
condensate band being small relative to the non-condensate band, and 
if $\Emax$ and $\ER$ are too large then these approximations will not be 
valid.  The approximations concerned are:
\Item[i)] That the non-condensate band is so large that it is
essentially undepleted by the process of condensate growth.
\Item[ii)] That the scattering processes between  condensate band atoms may be
taken as being negligible compared to the inter-band scattering processes, as
will be assumed later.   

It should be noted, and will be shown later, that the major effect on the
overall growth due to mean field effects is caused by the changes in energies
of the lowest energy levels, and for these levels the model proposed here is in
quite good agreement with the semiclassical results.  Because of this, as well
as for the above reasons, the value of $\ER$ used will generally remain
equal to $2\mu_C(n_{0,f})$.  The value of $\Emax$ is chosen somewhat 
larger.  This provides a check that the solutions we find 
do match smoothly onto the distribution above $\Emax$, which is 
assumed not to change.

\begin{figure}[t]
\epsfig{file=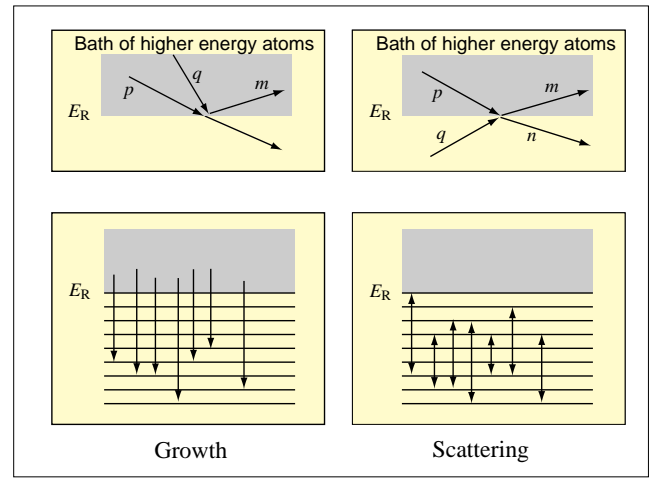,width=8.6cm}

\caption{Illustration of the two types of dynamical processes, growth 
and scattering.  Only the two non-condensate particle growth process 
is shown (process (a) in the text), and only 
non-condensate/condensate band scattering is shown (process (c)).}
\Label{fig:scattvsgrowth}
\end{figure}
\subsection{Dynamical Processes}
The dynamics which will be considered in order to describe the
evolution of the condensate band arise from the following processes:  
\Item[a)] Two particles in the non-condensate band collide, one of the
particles leaves with an increased energy, and the remaining particle enters
the condensate band, having now an energy less than $\Emax$.  Of course
the reverse process must also be considered---a non-condensate band particle
colliding with a condensate band particle and exciting it out of the condensate
band. 

\Item[b)] A non-condensate band particle collides with a condensate band
particle and exchanges energy such that both particles end up in the
condensate band, and the reverse process.

\Item[c)] A non-condensate band particle collides with a
condensate band particle, transferring some energy, but both particles
remain in their respective bands.

\Item[d)] Two particles in the condensate band collide, transferring energy,
with the result that both particles remain in the condensate band, but having
different energies than before the collision. 

\Label{sec:processes}

Processes (c) and (d) will be termed \emph{scattering processes}, since they do
not change the occupation number of either band.  Processes (a) and (b) cause
the number of particles in the condensate band to increase, and so will be
referred to as \emph{growth processes}.  The distinction between the two types
of processes is illustrated in Fig.\ref{fig:scattvsgrowth}. Because the
number of atoms in the non-condensate band is much larger than that in the
condensate band, the scattering will be dominated by processes of type (c), and
those of type (d) will be neglected.

These processes are described by the full quantum kinetic master equation
obtained in QKIII, which can be used to determine rate equations for the 
evolution of
the system.

\subsection{Growth Processes}
The formalism of \cite{trueBog,QK3,QK5} gives rise to rate equations for $ N$, 
the number of particles in the condensate band as a whole, and $ n_m$, which 
represent the number of quasiparticles in the $ m$th quasiparticle level. 
The derivation gives equations in the limit that $ N$ is sufficiently large for 
us to write $n_0 \approx N$.
The rate equations take the form:
\begin{eqnarray}\Label{Rate1}
{dn_m\over dt} &=& \left. \dot n_m\right|_{\rm growth}
+ \left. \dot n_m\right|_{\rm scatt}
\end{eqnarray}
The form of $ \left. \dot n_m\right|_{\rm growth} $ was given in QKIII, 
Sec.IV.E.3, and can be written in terms of the transition rates 
as follows
\begin{eqnarray}\Label{Rate2}
\dot n_m &=& \dot n^+_m +\dot n^-_m
\\ \Label{Rate3}
\dot N &=& 2W^+\left((1-e^{(\mu_C(N)-\mu)/k_{B}T})N+1\right)
\nonumber  \\ 
&&
+\sum_m\left\{ \dot n^+_m -\dot n^-_m\right\}
\end{eqnarray}
where
\begin{eqnarray}\Label{Rate4}
\dot n^+_m 
&\equiv& 2W^{++}_m(N)\left((1-e^{(\mu_C(N)-\mu+\epsilon_m)/k_{B}T})n_m+1\right)
\nonumber \\  \\
\Label{Rate5}
\dot n^-_m 
&\equiv& 2W^{-+}_m(N)\left((1-e^{(-\mu_C(N)+\mu+\epsilon_m)/k_{B}T})n_m+1
\right)
\nonumber \\
\end{eqnarray}
the relationships between forward and backward rates can be shown to
be
\begin{eqnarray}\Label{Rate6}
W^+(N) &=& e^{(\mu -\mu_C(N))/kT}W^-(N) \\
W_m^{++}(N) &=& e^{[\mu -\mu_C(N) -\epsilon_m]/k_{B}T}W_m^{--}(N)
\\  \Label{Rate7}
W_m^{+-}(N) &=& e^{[\mu -\mu_C(N) + \epsilon_m]/k_{B}T}W_m^{-+}(N).
\end{eqnarray}
In these equations, the energies of the quasiparticle excitations $\epsilon_m$
are measured from the \emph{ground state energy}, which is $\mu_C(N)$. 

\subsubsection{Transitions}
There are six  processes which are described by these equations:
\Item[i)] $N \rightarrow N + 1$, with no change in $\bn$.  The transition
probability for this process is $W^+(N) = R^+(\xi_N,\mu_C(N)/\hbar)$.

\Item[ii)] $N \rightarrow N - 1$, with no change in $\bn$.  The transition
probability for this process is $W^-(N) = R^-(\xi_{N-1},\mu_{N-1}/\hbar)$.

\Item[iii)] $N \rightarrow N + 1$, with $n_m \rightarrow n_m+1$.  The 
transition
probability for this process is $W_m^{++}(N) =
R^+(f_m,(\epsilon^m_N+\mu_C(N))/\hbar)$.

\Item[iv)] $N \rightarrow N - 1$, with $n_m \rightarrow n_m-1$.  The transition
probability for this process is $W_m^{--}(N) =
R^-(f_m,(\epsilon^m_{N-1}+\mu_{N-1})/\hbar)$.

\Item[v)] $N \rightarrow N + 1$, with $n_m \rightarrow n_m-1$.  The transition
probability for this process is $W_m^{+-}(N) =
R^+(g_m,(-\epsilon^m_{N}+\mu_{N})/\hbar)$.

\Item[vi)] $N \rightarrow N - 1$, with $n_m \rightarrow n_m+1$.  The transition
probability for this process is $W_m^{-+}(N) =
R^-(g_m,(-\epsilon^m_{N-1}+\mu_{N-1})/\hbar)$.


Here the functions $f_m$ and $g_m$ are amplitudes for the creation 
and destruction of atoms in quasiparticle states with energies 
$\epsilon^m_N$, which are defined in~\cite{trueBog,QK3}, but whose 
explicit form will not need to be used here; $\xi_N$ is the 
condensate wavefunction for $N$ atoms.  The functions 
$R^\pm(y,\omega')$ are defined by \widetext
\begin{eqnarray}
 R^{+}(y,\omega')&=&{u^2\over(2\pi)^5\hbar^2}\int d^3\br
\int d^3{\bf K}_1 d^3{\bf K}_2 d^3{\bf K}_3 d^3{\bf k}\, 
\delta(\Delta\omega_{123}(\br)-\omega')  
\delta({\bf K}_1 + {\bf K}_2 - {\bf K}_3 - {\bf k})
F_1F_2 (1+F_3) {\cal W}_y(\br,{\bf k}) \Label{Rplus}
 \\ 
 R^{-}(y,\omega')&=&{u^2\over(2\pi)^5\hbar^2}\int d^3\br
\int d^3{\bf K}_1 d^3{\bf K}_2 d^3{\bf K}_3 d^3{\bf k}\,
\delta(\Delta\omega_{123}(\br)-\omega')  
\delta({\bf K}_1 + {\bf K}_2 - {\bf K}_3 - {\bf k})
( 1+F_1)( 1+F_2)F_3 {\cal W}_y(\br,{\bf k}) 
\nonumber \\
\end{eqnarray}
\narrowtext\noindent
In these equations the following notation is used
\begin{eqnarray}
\hbar\omega_{\bK_i}(\br) &\equiv& {\hbar^2 \bK_i^2 \over 2m} + V_T(\br),\\ 
\Delta\omega_{123} &\equiv& \omega_{\bK_1} +\omega_{\bK_2} - \omega_{\bK_3},\\
\Delta\bK &\equiv& {\bf K}_1 + {\bf K}_2 - {\bf K}_3 - {\bf k}, \\
u&=& {4\pi\hbar^2 a \over m}.
\end{eqnarray}
 
The function $F_i = F(\bK_i,\br)$ is the distribution function for 
the non-condensate particles, and the Wigner function is given by
\begin{eqnarray}
 {\cal W}_y(\br,{\bf k}) = {1\over (2\pi)^3}\int d^3{\bf v}\,
y^*\left(\br+{{\bf v}\over 2}\right)y\left(\br-{{\bf v}\over 2}\right)
e^{i{\bf k}\cdot{\bf v}} .
\end{eqnarray}
The $R^+$ function is related to collisions between two non-condensate band
particles with momenta $\bK_1$ and $\bK_2$, from which the particles leave with
momenta $\bK_3$ and ${\bf k}$.  The particle with momentum ${\bf k}$ is now in
the condensate band (ie: ${\bf k}$ is small) so that the particle with momentum
$\bK_3$ remains in the non-condensate band.  The functions $R^\pm(y,\omega')$
therefore represent the rates for collisions which result in a particle
entering ($+$) or leaving ($-$) the condensate band with an energy
$\hbar\omega'$.

\subsubsection{Approximate evaluation of transition rates}
The rate factors $W^{\pm}(N)$ contain integrals over all space of terms
containing the product of $F(\bK,\br)$ terms times the Wigner function
corresponding to the ground state wavefunction ${\cal W}_{\xi_N}(\br,\bK)$.  In
practice the ground state wavefunction is very sharply peaked in comparison to
the spatial dependence of the $F(\bK,\br)$ functions which describe the
remainder of the cloud of atoms.  This means that in the spatial integral for
the $W^{\pm}(\br)$ terms (from equation~(\ref{Rplus})) the $F(\bK,\br)$ terms
can be approximated by their values at $\br =0$.  This gives  
\widetext
\begin{eqnarray}\Label{simp2}
W^+(N)&=& {u^2\over(2\pi)^5\hbar^2}
\int d^3{\bf K}_1\int d^3{\bf K}_2\int d^3{\bf K}_3\int d^3{\bf k}\,
\delta(\Delta\bK)
\delta(\Delta\omega_{123}({\bf 0})-\mu_C(N)/\hbar)
F({\bf K}_1,{\bf 0})F({\bf K}_2,{\bf 0}) \nonumber\\
&&\quad\quad\quad\times \bigl( 1+F({\bf K}_3,{\bf 0})\bigr)
|\tilde \xi_N({\bf k})|^2
\Label{W+simp} \\ 
W^-(N+1)&=&
 {u^2\over(2\pi)^5\hbar^2}
\int d^3{\bf K}_1\int d^3{\bf K}_2\int d^3{\bf K}_3\int d^3{\bf k}\,
\delta(\Delta\bK)
\delta(\Delta\omega_{123}({\bf 0})-\mu_C(N)/\hbar)
\bigl( 1+F({\bf K}_1,{\bf 0})\bigr) \nonumber\\ 
&&\quad\quad\quad\times\bigl( 1+F({\bf K}_2,{\bf 0})\bigr)
F({\bf K}_3,{\bf 0})
|\tilde \xi_N({\bf k})|^2 \Label{W-simp} 
\end{eqnarray}
\narrowtext\noindent
in which $ \tilde \xi_N({\bf k})$ is the momentum-space ground-state
wavefunction, obtained from the spatial form by
\begin{eqnarray}\Label{simp4}
\tilde \xi_N({\bf k}) &=& {1\over(2\pi)^{3/2}}\int d^3{\bf r}\,
e^{i{\bf k}\cdot{\bf r}}\xi_N({\bf r})
\end{eqnarray}

In~\cite{BosGro}, progress was made by assuming that the non-condensate band
distribution $F(\bK,\br)$ was given by the classical Maxwell-Boltzmann
distribution
\begin{equation}
F(\bK,\br) \approx {\exp{\left(- {{\hbar^2\bK^2/2m}+V_T(\br) -\mu
\over {kT}}\right)}}.
\end{equation}
with values of $T$ and $\mu$ which ensure the formation of a condensate once
the system reaches equilibrium.  Furthermore, in calculating the integrals in
equations~(\ref{W+simp}) and~(\ref{W-simp}), it was assumed that the range of
condensate band energies was negligible compared to that of the non-condensate
band.  Thus the range of ${\bf k}$ was negligible compared to the range of
$\bK_{1,2,3}$ and the integrals in $\bK_{1,2,3}$ were calculated over all
energies rather than just over the non-condensate band.  The function
$F(\bK,\br)$ was also assumed to be negligible compared to unity. The result
obtained was~\cite{BosGro}
\begin{equation} \Label{wplusMB}
W^+(N) = {4m(akT)^2 \over \pi\hbar^3}e^{2\mu/kT}
\left[ {\mu_C(N) \over kT} K_1\left( {\mu_C(N) \over kT} \right) \right].
\end{equation} 
Here $K_1(x)$ is a modified Bessel function.  In almost all practical
situations the term in square brackets in the above equation is approximately
equal to unity, and so $W^+$ is essentially independent of $N$.  The value of
$W^-(N)$, the rate of transitions out of the condensate band, can be obtained 
in
a similar fashion to that for $W^+$.  The ratio of forward to backward rates
is found to be given by
\begin{equation} \Label{wplusvswminus}
W^+(N) = e^{(\mu -\mu_C(N))/kT}W^-(N),
\end{equation}
which stems from the assumption of the thermal undepleted bath, and the
definitions of $W^+$ and $W^-$.  From this equation and the master equation it
can be seen that equilibrium is achieved when $\mu_C(N) = \mu$ to order $1/N$.

\subsubsection{Simple growth equation}
A rate equation for the mean number of atoms in
the condensate $\langle N \rangle$ (written as $N$ for convenience for the rest
of this section) was obtained in \cite{BosGro,QK3} 
\begin{equation} \Label{simplegrowth}
\dot N = 2W^+(N)\left\{\left(1-e^{(\mu_C(N)-\mu)/k_{B}T}\right)N +1\right\}.
\end{equation}
This equation is the \emph{simple growth equation} used for the simulations of
condensate growth in~\cite{BosGro,QK3}.  If the system starts with $N=0$, the
growth begins slowly (but at a finite rate).  Once a significant condensate
occupation is attained the term proportional to $N$ becomes dominant, causing a
much faster growth rate.  The growth eventually slows as $\mu_C(N)$ approaches
$\mu$ and the system settles into equilibrium.  This gives the curve an
$S$-shape as will be shown in the following section.

By using the Thomas-Fermi chemical potential (\ref{TFmu2}) for
$\mu_C(N)$, and the Maxwell-Boltzmann form for $W^\pm(N)$, the first 
simulations
for the growth of a realistic condensate were presented in~\cite{BosGro}.  The
growth equation is simple to solve numerically, for whatever number of
particles is necessary (for example, the growth of a condensate containing 5
million atoms was simulated in~\cite{BosGro}). 

\subsubsection{Beyond the simple growth equation}
The derivation of the simple growth equation contains a number of
approximations and simplifications.  The major behavior once $N$ becomes large
should be described quite well by the simple growth equation, but terms which
were neglected may have significant effects during the initial stages of
growth. Possibly significant factors which should first be considered are
\Item[i)] The effect of considering all quasiparticle levels (the excited 
levels
in the condensate band).
\Item[ii)] The effect of scattering processes (as defined in
section~\ref{sec:processes}).
\Item[iii)] Corrections to the $W^\pm(N)$ terms to consider the more realistic
Bose-Einstein distribution function.
\Item[iv)] The fluctuations around the mean number.
%

\Item[]This paper aims to consider the effects of incorporating the first three 
of these factors into the growth equation.  
During the process of BEC formation, the spectrum of eigenvalues makes 
a transition from the unperturbed spectrum of trap levels to the case 
where the spectrum is strongly affected by the condensate in the 
ground state.  The Bogoliubov spectrum of a condensed gas is valid in 
the case where the number of particles in the condensate, $ n_0$, is 
so large that it is valid to write $ n_0 \approx N$.  Thus, during the 
initial stages of condensate formation, where this is not true, one 
must use another formalism.  In this paper we will consider the 
situation in which the interaction between the particles is very weak, 
as is in practice the case.  This means that we will be able to use 
the unperturbed spectrum for the initial stages of condensation, and 
only use the Bogoliubov description once enough condensate has formed 
to make the effective interaction rather stronger.

The basic formalism of \cite{QK3} can still be carried out in this 
case, and the modification that is found is rather 
minor---essentially, we make the substitution $ N\to n_0$ in the 
chemical potential and the $ W^+(N), W^{++}(N)$ functions, and set 
$W^{-+}_m\to 0 $, since this term comes from the mixing of creation 
and annihilation operators which arises from the Bogoliubov method.  

By making these adjustments, and now grouping the levels into sub-bands of
energy $e_m$ (measured now from zero, rather than from $\mu_C(N)$ as was the 
case for $\epsilon_m$), with each sub-band containing $g_m$ levels, the 
equations of motion for the growth processes are now
\begin{eqnarray}\Label{growth}
\dot n_m|_{\rm growth} 
&=& 
2W_m^{++}(n_0)\left\{\left[1-e^{e_m-\mu \over kT}\right]n_m+g_m
\right\}, 
\\ \Label{grow0}
\dot n_0|_{\rm growth}  &=& 
2W^+(n_0)\left\{\left[1-e^{\mu_C(n_0)-\mu\over kT}\right]n_0+1
\right\}. 
\end{eqnarray}
We will make the further---possibly rather drastic---simplification, and 
neglect entirely the effect of phonon-like quasiparticles, which are known to 
comprise only a very small fraction of the levels normally occupied at the 
temperatures considered. Thus the
excited states are now taken to be of a purely
single-particle nature, and the condensate band is now described by the
occupation number of the condensate level (the lowest energy level) $n_0$, and
by the occupation numbers of each of the excited states $n_m$.   In this case, 
the equations (\ref{growth},\ref{scatt}) become the same as the 
equations
(\ref{Rate1}--\ref{Rate7}), and may therefore be used to represent the full 
condensate growth process.

\Label{sec:growth}

\subsubsection{Evaluation of Transition Probabilities}
The value for the transition probability $W^+(N)$ found in the simple growth
equation   (\ref{simplegrowth}) was derived
by making some rather sweeping assumptions, and as such
equation~(\ref{wplusMB}) is really just an order of magnitude estimate.  To
obtain a more accurate value the full Bose-Einstein distribution must be used
for $F(\bK,\br)$ and the ranges of integration of the non-condensate functions
must exclude the condensate band in which $F(\bK,\br)$ would become very
large.  We then have from (\ref{W+simp})
\widetext
\begin{eqnarray}
W^+(n_0)&=& {u^2\over(2\pi)^5\hbar^2} \int d^3\br
\int\limits_{E > \Emax}{\!\!\!\!\!\!\!} d^3{\bf K}_1
\int\limits_{E > \Emax}{\!\!\!\!\!\!\!} d^3{\bf K}_2
\int\limits_{E > \Emax}{\!\!\!\!\!\!\!} d^3{\bf K}_3
\int\limits_{E < \Emax}{\!\!\!\!\!\!\!} d^3{\bf k}\,
\delta(\Delta\bK)\delta(\Delta\omega_{123}({\bf r})-\mu_C(n_0)/\hbar)
F({\bf K}_1,{\bf r}) \nonumber\\
&&\quad\quad\quad\times F({\bf K}_2,{\bf r}) 
\bigl( 1+F({\bf K}_3,{\bf r})\bigr)
{\cal W}_{\xi_{n_0}}(\br,{\bf k})
\end{eqnarray}
\narrowtext\noindent
with
\begin{eqnarray}
F(\bK,\br) &=& \left[\exp{\left({\hbar^2\bK^2 / 2m}+V_T(\br) -\mu
 \over {kT}\right)}-1\right]^{-1}  \\
&& = \sum_{s=1}^\infty \exp{\left(-s\left[{\hbar^2\bK^2 / 2m}+V_T(\br) -\mu
 \over {kT}\right] \right)}.
\end{eqnarray}

Again we have to make the approximations:
\Item[i)] That the spatial dependence can be neglected, so $F(\bK,\br) 
\rightarrow
F(\bK,{\bf 0})$ and ${\cal W}_{\xi_N}(\br,{\bf k}) \rightarrow 
|\xi_N({\bf k})|^2$.
\Item[ii)] That we can neglect the ${\bf k}$ dependence, except in $\xi_N$, so 
that
the only ${\bf k}$ dependence left is removed by $\int d^3{\bf k} |\xi_N({\bf
k})|^2 = 1$.
%
The integrals over $\bK_1$, $\bK_2$ and $\bK_3$ can then be performed, to give 
a final form for $W^+(n_0)$, found by Davis~\cite{Mattcomm}, of
\begin{eqnarray}
W^+(n_0) &=& {1 \over 2}\left({k_BT\over\hbar\omega}\right)^2
\bigg\{[\log(1- z)]^2 
\nonumber \\
&& +  z^2\sum_{r=1}^{\infty}[ z\, z(n_0)]^r
[\Phi( z,1,r+1)]^2\bigg\},  \Label{wplus}
\end{eqnarray}
where
\begin{eqnarray}\Label{newWplus2}
 z &=& e^{\left(\mu-\Emax\over k_BT\right)} 
\ \ \ \ z(n_0) = e^{\left(\mu_C(n_0)-\Emax\over k_BT\right)} 
\Label{newWplus3}.
\end{eqnarray}
The function $ \Phi$ is the {\em Lerch transcendent} \cite{Bateman}, defined by
\begin{equation}
\Phi(x,s,a) = \sum_{k=0}^\infty {x^k/ (a+k)^s}. \Label{lerch}
\end{equation}

This form of $W^+(n_0)$ gives values of about a factor of three greater than
the previous form in equation~(\ref{wplusMB}), depending on the exact 
parameters
of the system, and this gives a correspondingly faster growth than that
in~\cite{BosGro}. 

The values for $W_m^{++}(n_0)$ are more difficult to obtain.  The
$W_m^{++}(n_0)$ terms are the average of the $W^{++}$ terms for all the
individual levels in the sub-band.  The $W^{++}$ terms are given by similar
overlap integrals as used for the $W^+$ terms, and for the lower energy levels
in the condensate band, the overlap of the wavefunction with the spatial
distribution of a non-condensate band particle should be similar to that for
the condensate level.  Thus it is expected that the $W^{++}$ terms should be 
of the same order of magnitude as $W^+(n_0)$.  Progress can therefore be made
by approximating $W_m^{++}(n_0) \approx W^{+}(n_0)$.  The effect and validity
of this approximation will be investigated in the following section.

\subsection{Scattering Processes}
\Label{ScattProc}%
Scattering processes in this paper also need to be included in the 
evolution of
$n_m$. 
Scattering between two atoms in the non-condensate band does not have to be
explicitly considered, since it has been dealt with in making the assumption
that the non-condensate band is an equilibrated time-independent thermal bath. 
Furthermore, the scattering between two condensate band atoms will be neglected
since, at any time, the number of atoms in the condensate band is small
relative to the number in the non-condensate band.  The dominant scattering
processes, and the only ones which will be considered, are the scattering of
atoms between levels in the condensate band, due to interactions with
non-condensate band atoms (see Fig.\ref{fig:scattvsgrowth}).  These stem
from terms in the full master equation which involve two condensate field
operators $\phi$.  These terms give rise to a master equation of the 
form, as shown in QKIII equation (50d):
\begin{eqnarray} 
&& \dot\rho |_{\rm scatt} = \!\!\sum_{mk\atop\epsilon_k<\epsilon_m}\!\!
\gamma_{km}\bar N_{km} \left\{2 X_{km}\rho
X^\dagger_{km}-  [ X^\dagger_{km} X_{km},\rho]_{+}\right\} 
\nonumber\\
&&\quad
+
\!\!\sum_{mk\atop \epsilon_k>\epsilon_m}\!\!
\gamma_{km}(\bar N_{km}+1) \left\{2
X_{km}\rho X^\dagger_{km}-  [ X^\dagger_{km} X_{km},\rho]_{+}\right\}
\nonumber\\ 
&&\quad +
\!\!\sum_{km\atop \epsilon_k=\epsilon_m}\!\!
\gamma_{km}\bar M_{km}
\left\{2 X_{km}\rho X^\dagger_{km}-  [ X^\dagger_{km} X_{km},\rho]_{+}\right\}
\nonumber\\ \Label{scattmaster} 
\end{eqnarray} 
which is equivalent to the master equation governing the scattering of
particles by a heat bath~\cite{QN}.

Here the operators are defined
by 
\begin{eqnarray} 
X_{km}&\equiv & a^\dagger_ma_k 
\end{eqnarray} 
where $
a_{k}$ is the destruction operator for an atom in state $k$ with energy
$\epsilon_k$.  As in the previous section, we treat all excitations as being 
particle-like.  The rates of the processes are determined by the factors
$\gamma_{km}$, and the factors $\bar N_{km}$ are defined by 
\begin{eqnarray}
\bar N_{km} \equiv {1\over \exp\left(\epsilon_k-\epsilon_m\over kT\right)-1}.
\end{eqnarray}

The last line in the master equation~(\ref{scattmaster}) represents scattering
between degenerate energy levels, which will not have any contribution to the
time dependence of $n_m$ once the levels are grouped into sub-bands, and so
can be ignored.

The corresponding rate equation for $n_k = \langle a_k^\dagger a_k\rangle$, the
mean occupation of the $k$th level, can easily be found from the master
equation. 
When levels are grouped into sub-bands with mean energy $e_k$, occupation
$n_k$, and with $g_k$ levels contained in the sub-bands, it
becomes
\begin{eqnarray} && \dot n_k|_{\rm scatt} = \nonumber\\ && \sum_{m\atop
e_k>e_m}\gamma_{km} \left\{\bar N_{km}n_m(n_k+g_k) -(\bar
N_{km}+1)(n_m+g_m)n_k\right\} \nonumber \\  && + \sum_{l\atop
e_k<e_l}\gamma_{lk} \left\{(\bar N_{lk}+1)n_l(n_k+g_k) -\bar
N_{lk}(n_l+g_l)n_k\right\}. 
\nonumber \\
\Label{scattqk} 
\end{eqnarray} 
The transition rates
$\gamma_{km}$ now represent averages over the all the individual level
transition rates which transfer an atom from the $k$ sub-band to the $m$
sub-band. 
The transition probabilities $\gamma_{km}$ can be found in a similar manner to
the $W^+(n_0)$ terms, from terms of the form~\cite{QK3}
\widetext
\begin{eqnarray}
R_{km}(N) &=&{4u^2\over(2\pi)^5\hbar^2}\int d^3\br
\int d^3{\bf K}_1\int d^3{\bf K}_2\int d^3{\bf k}\int d^3{\bf k}'
\delta({\bf K}_1 - {\bf K}_2 - {\bf k} + {\bf k}')
F({\bf K}_1,\br)\bigl( 1+F({\bf K}_2,\br)\bigr)
\nonumber \\
&& \qquad\qquad\times
{\cal W}_k(N,\br,{\bf k}){\cal W}_m(N,\br,{\bf k}') 
\delta(\Delta\omega_{12}(\br)-\Omega_m+\Omega_k),
\end{eqnarray}
\narrowtext\noindent
where 
\begin{eqnarray}
\Omega_m(N) &=& {\mu_C(N) + \epsilon_m(N)\over\hbar}
\\
&=& {e_m(N)\over\hbar}
\end{eqnarray}
and the rest of 
the
notation is as was used in the previous section. 
\subsubsection{Estimates for $\gamma_{km}$}
Explicit  computation
involved in calculating these factors is impractical, and, it will turn out, 
unnecessary when the scattering is sufficiently strong.
We shall instead estimate these $\gamma_{km}$ rates by using the quantum 
Boltzmann approach of 
Holland \emph{et al.}~\cite{HWC1997}.

By treating the excitation spectrum as given by the eigenstates of the trapping
potential, without modification by the presence of the condensate, and by using
the ergodic assumption  Holland \emph{et al.}
obtained the kinetic equation (equation (12) in~\cite{HWC1997})
\begin{eqnarray}
g_n{\partial f_n \over\partial \tau} &=&\sum_{mqp}\delta_{e_n+e_m,e_p+e_q}
g(e_m,e_n,e_p,e_q) 
\\ \nonumber &\times&\bigg[ f_qf_p(1+f_m)(1+f_n) -(1+f_q)(1+f_p)f_mf_n\bigg],
\end{eqnarray}
in which $\tau ={ (8ma^2\omega^2/\pi\hbar)} t $.  The population of a level
with energy $e_n$ is $f_n$, and the degeneracy of levels at that energy is
$g_{n}$.  The collision kernel $g(e_m,e_n,e_p,e_q)$ is given from the overlap
integrals for the states $m,n,q$ and $p$.  However, when the energies of the
levels considered are spread quite far apart, Holland \emph{et al.} found from
numerical calculations that the collision kernel $g(e_{\rm min},e_n,e_p,e_q)$
is well approximated by the degeneracy $g_{e_{\rm min}}$. Here $e_{\rm min}$ is
the smallest of the energies in the collision.  In our model the energies will
always be quite well spread, as: a) they must be in different sub-bands, each
sub-band being quite well separated from the next in terms of mean energy; and
b) the scattering processes we are attempting to describe must have $e_m$ and
$e_p$ in the non-condensate band and $e_q$ and $e_p$ in the condensate band. 
Thus in our model we may safely use $g(e_{\rm min},e_n,e_p,e_q) \approx
g_{e_{\rm min}}$.

By summing over $m$ and $p$ terms (which are levels higher than $\Emax$),
the effect of all the non-condensate (`bath') levels on the condensate band
atoms may be calculated.  The kinetic equation for the scattering now becomes
\begin{eqnarray}
&&\left.{\partial n_n \over\partial\tau }\right |_{\rm scatt} =
\nonumber\\&&\,
\sum_q\Bigg[(1+f_n)f_q
\left(\sum_{mp}\delta(n,m;p,q)g_{e_{\rm min}} f_p(1+f_m)\right)
\nonumber \\ &&\,
 - (1+f_q)f_n
\left(\sum_{mp}\delta(n,m;p,q)g_{e_{\rm min}} f_m(1+f_p)\right)
\Bigg] \Label{kineq1}
\end{eqnarray}
where the following notation has been used
\begin{eqnarray}
n_m&=& f_m g_m
\\ 
\delta(n,m;p,q) &\equiv& \delta_{e_n+e_m,e_p+e_q}.
\end{eqnarray}
The terms of this equation can be simplified for the different possible cases.

\noindent {\em First Line, case $e_n > e_q$:}
In this case, $e_q = e_{\rm min}$, and energy conservation is satisfied when
\begin{equation}
e_p = e_m + \hbar\omega_{nq}
\end{equation}
where $\hbar\omega_{nq} = e_n - e_q$.  The summation term in the first line of
equation~(\ref{kineq1}) then becomes
\begin{eqnarray}
g_q\sum_{e_m>\Emax}(1+f_m)f_{m+\omega_{nq}}&\approx&
 g_qe^{(\mu-\hbar\omega_{nq})/kT}\Gamma(T) 
\\
\mbox{where }\Gamma(T) &\equiv&\sum_{e_m>\Emax}e^{-e_m/kT} 
\Label{Gammadef}
\end{eqnarray}
The approximation which has been made is that $(1+f_m) \approx 1$, which should
be acceptable since state $m$ is of high energy (ie: in the non-condensate
band).  The calculation of $\Gamma(T)$ requires knowledge of the spectrum of
energies in the non-condensate band, which is complicated for an anisotropic
trap.  The form for an \emph{isotropic} harmonic potential $\bar{\Gamma}(T)$ is
easily calculated though, and gives
\begin{eqnarray}
\bar{\Gamma}(T) ={ e^{-\Emax/kT}\over1-e^{-\hbar\omega/kT}} 
\Label{Gamma}
\end{eqnarray}
where $\omega$ is the frequency of the potential.  We will therefore make the
approximation that $\Gamma(T) \approx \bar{\Gamma}(T)$ for the determination of
$\gamma_{km}$, using the geometrical mean frequency of the real trap as the
frequency of the isotropic potential.  Thus $\omega =
(\omega_x\omega_y\omega_z)^{1/3}$ in equation~(\ref{Gamma}).

\noindent {\em Remaining terms:}
Similar reasoning leads to the results
\begin{eqnarray}
&&g_n\sum_{e_p>\Emax}(1+f_{p+\hbar\omega_{qn}})f_p
\approx g_n\sum_{e_p>\Emax}e^{(\mu-e_p)/kT}
\\&&\qquad\qquad\qquad\qquad\qquad\quad
 =  g_ne^{\mu/kT}\Gamma(T) \\
&&g_q\sum_{e_m>\Emax}f_m(1+f_{m+\hbar\omega_{nq}}) 
\approx g_qe^{\mu/kT}\Gamma(T,R) \\
&&g_n\sum_{e_p>\Emax} f_{p+\hbar\omega_{qn}}(1+f_p) 
\approx g_ne^{(\mu-\hbar\omega_{qn})/kT}
\Gamma(T)
\end{eqnarray}

\subsubsection{Total Scattering Equation:}
The total kinetic equation governing the scattering processes is now given by
\begin{eqnarray}
&&\left.{\dot n_m }\right |_{\rm scatt}
= {8ma^2\omega^2 \over \pi \hbar}
e^{\mu/k_BT}\Gamma(T)\times
\nonumber\\ 
&&\,\,
\Bigg\{\sum_{k<m}{1\over g_m}\left[n_k(g_m+n_m)
{e^{-\hbar\omega_{mk}/k_BT} }
-n_m(g_k+n_k)\right]
\nonumber \\
 &&\,\,  +
\sum_{k>m}{1\over g_k }\left[n_k(g_m+n_m)
-n_m(g_k+n_k){e^{-\hbar\omega_{km}/k_BT}}\right]\Bigg\}.
\nonumber \\ \Label{scatt}
\end{eqnarray}
The notation $k>m$ is now being used to mean $e_k > e_m$. This is the rate
equation governing scattering processes, it is equivalent to equation
(\ref{scattqk}) if the following transformations are made
\begin{eqnarray}
\bar N(\omega_{nq}) & \to & e^{-\hbar\omega_{nq}/kT}
\\ 1 +\bar N(\omega_{nq}) & \to & 1
\\ \gamma_{nq}&\to &{ 8ma^2\omega^2 \over \pi \hbar}
{ e^{\mu/kT}\Gamma(T)\over g_n} \qquad\mbox{when }n>q
\\ \gamma_{nq}&\to & {8ma^2\omega^2 \over \pi \hbar}
{e^{\mu/kT}\Gamma(T)\over g_q} \qquad\mbox{when }n<q
\end{eqnarray}

\subsection{Rate equations including scattering and growth}
The total rate equation governing the evolution of this system is then given by
adding equation~(\ref{growth}) to equation~(\ref{scatt})
\begin{eqnarray}
\dot n_{m} &=& \dot n_{m}|_{\rm growth} + \dot n_{m}|_{\rm scatt}
\Label{totalgrowth}
\end{eqnarray}
and for the condensate level evolution (\ref{grow0}) is used in
place of equation~(\ref{growth}).

It is useful here to review the major approximations that have been made in
order to derive these equations.  It was assumed that

\Item[i)] The non-condensate band is very large, so it is essentially 
undepleted,
and it is in equilibrium.

\Item[ii)] The influence of collective excitations is negligibly small, so that 
the
states in the condensate band are all of single-particle nature.

\Item[iii)] The density of states in the system is as described in
section~\ref{sec:densstates}.

\Item[iv)] The ergodic approximation is valid, and that states in the
condensate band which have similar energies may be `binned' together for the
purpose of describing their evolution.

\Item[v)] The fluctuations of the occupation numbers around the mean numbers 
may be
ignored.

\Item[vi)] The rate constants $W^{++}$ for the growth processes which change
the occupations of the excited states in the condensate band are equal to the
rate constant for the growth of the condensate level.

\Item[vii)] The rate constant for the scattering processes in an anisotropic 
well
of geometrical mean frequency $\omega$ is equal to that in an isotropic well
with frequency $\omega$.

\Item[viii)] Three body collisional processes may be ignored.
%
%

\section{Numerical Solutions of the Growth}
The rate equations derived to describe the growth of a condensate in 
the previous section are quite straightforward to solve numerically, 
and the solutions can be obtained in a matter of a few seconds as 
opposed to other numerical solutions which have been very time 
consuming.  The nature of these solutions will be discussed and 
comparisons will be made with experimental data published in 
\cite{MITgrowth}.  The parameters of the system modelled were chosen 
to be the same as in the MIT growth experiments, that is: 
\Item[i)]  Using a dilute gas of $^{23}{\rm Na}$ atoms, characterised by an 
$ s$-wave
scattering length of $a = 2.75{\rm nm}$~\cite{scattlength,MITgrowth}.
\Item[ii)]  With an axially symmetric (`cigar shaped') harmonic trapping 
potential
described by the frequencies $\omega_x=\omega_y = 2\pi \times 82.3{\rm Hz}$ and
$\omega_z = 2\pi \times 18{\rm Hz}$, and giving a geometrical mean frequency of
$\omega = 2\pi \times 50{\rm Hz}$~\cite{MITgrowth,MITpersonal}.
\Item[iii)]  With a total number of atoms of the order of $10^7$.  The numbers 
found
in the condensate level once equilibrium is reached are in the range $5\times
10^5$ to $1 \times 10^7$, giving a condensate occupation of between 5 and 30\%
of the total number of atoms~\cite{MITgrowth,MITpersonal}.  In the majority of
cases the thermal bath was therefore depleted by only a small amount and the
undepleted model which is used here should be a good approximation.  
\Item[iv)]  With temperatures in the range $0.5\mu{\rm K}$ to $1.5\mu{\rm
K}$~\cite{MITgrowth,MITpersonal}.

\subsection{Results}
\subsubsection{Simple Growth Equation Results}

Numerical solutions of the simple growth equation
\begin{equation} \Label{simplegrowth2}
\dot n_0 = 2W^+(n_0)\left\{\left(1-e^{(\mu_C(n_0)-\mu)/k_{B}T}\right)n_0 +1
\right\},
\end{equation}
are easily obtained \cite{BosGro}, and an example of the resulting 
growth curve is shown in Fig.\ref{fig:simplegrowth}.  This curve shows 
a characteristic S-shape, the slow initial growth occurs as a result 
of spontaneous ($\mbox{}+1$) terms in equation~(\ref{simplegrowth2}) 
and then, once the occupation becomes large enough, the stimulated 
growth terms (those proportional to $n_0$) dominate and the growth 
accelerates.  The condensate grows quickly, until it approaches 
equilibrium where it slows again as $\mu_C(n_0)$ approaches $\mu$, 
giving the final part of the S-shape nature.

\begin{figure}
\epsfig{file=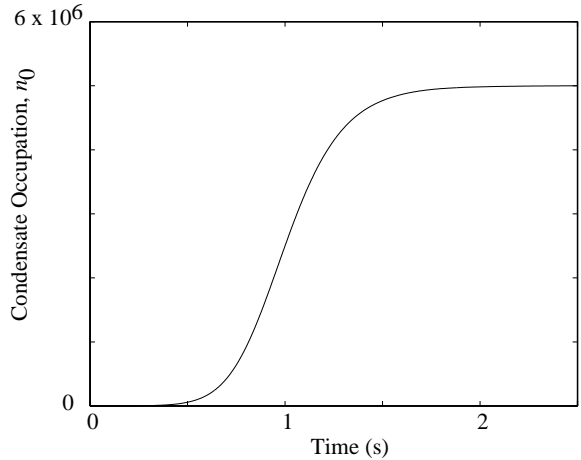,width=8cm}
\caption{Typical results of the simple growth equation for
the growth of a Bose-Einstein Condensate in the MIT apparatus at a temperature
of $900{\rm nK}$ with $\mu = 43.3\hbar\omega$.  The initial condensate
occupation was taken to be 100 atoms.}
\Label{fig:simplegrowth}
\end{figure}

The derivation of the simple growth equation contained a number of significant
approximations and assumptions.  In effect it describes the situation in which 
the occupations of all levels higher in energy than the ground state are
treated as time independent, giving an undepleted thermal bath in contact with
only the condensate level.  Furthermore, the populations of the levels in
this bath are given by the Maxwell-Boltzmann distribution, rather than the
correct Bose-Einstein distribution, and no mean-field effects are introduced. 
The simple growth equation is therefore just a first approximation for the
growth, and was merely intended to give an order of magnitude 
description of the growth process. 
Indeed, as shall be shown a subsequent section, experimental
measurements~\cite{MITgrowth} have shown that, although it gives the correct
order of magnitude for timescale of the growth, the simple growth equation does
not describe the rate of growth to any closer than a factor of about 3.

\subsubsection{New Model of Growth Processes}

In order to improve the description of the growth over that of the simple
growth equation, the first improvement which will be made is the more accurate
calculation of $W^+$, using the full Bose-Einstein distribution.  However, for
lower energy levels the equilibrium populations determined by the Bose-Einstein
distribution are very large.  Having such large populations in these levels
will obviously not be a good model of a system rapidly cooled from a point
where $\mu$ was negative to a region where $\mu$ has become positive, since the
changes in populations of these levels required during this process are so
substantial that a fairly long time will be required for the levels to come to
equilibrium.  It is therefore unphysical to consider a situation where the low
energy levels have reached their equilibrium populations before the condensate
level has even started to evolve.  For these reasons, it is not consistent to
simply use the Bose-Einstein distribution to find $W^+$ in the model considered
by the simple growth equation.

In order to develop a consistent description there need to be a number of lower
energy levels with time-dependent populations considered, forming the
condensate band.  Furthermore, the energies of these levels will be increased
by the growth of the condensate, due to increased mean field interactions.  To
describe this situation the model described in the previous section must be
implemented.  Considering at first only the `growth' processes, the evolution
was found in section~\ref{sec:growth} to be given by 
\begin{eqnarray} 
\dot
n_m|_{\rm growth}  &=&  2W_m^{++}(n_0)\left\{\left[1-e^{e_m-\mu \over
kT}\right]n_m+g_m \right\}, 
\Label{growth2} \\  
\dot n_0|_{\rm growth}  &=& 
2W^+(n_0)\left\{\left[1-e^{\mu_C(n_0)-\mu\over kT}\right]n_0+1 \right\}.
\Label{gsgrowth2b} 
\end{eqnarray} 
where, for reasons given in the previous
section, we approximate $W_m^{++}(n_0)$ by $W^+(n_0)$.
This form of evolution is essentially the simple growth
equation applied to several energy levels.

It is now possible to perform a more accurate calculation of $W^+$,
using the Bose-Einstein distribution to describe the population of levels above
$\Emax$ (ie: in the non-condensate band), and summing only over the
non-condensate band levels.  The new form of $W^+$ is given by
equation~(\ref{wplus}).  

Sample solutions to the coupled differential equations~(\ref{growth2})
and~(\ref{gsgrowth2b}) are shown in Fig.\ref{fig:samplenewWplus}.  The rate
of growth of the condensate has increased substantially (generally by at least
a factor of 3) over that predicted by the simple growth equation, this is due 
to
the more accurate calculation of $W^+(n_0)$. However, the shape of the growth 
is
still essentially the same as that given by the simple growth equation.

It can be seen from Fig.\ref{fig:samplenewWplus} that the lower energy
levels also experience very substantial growth in this model.  Indeed the
occupations of some of these levels can exceed the condensate occupation
substantially before relaxing back to their equilibrium values.  This 
is of course not a realistic scenario, and is certainly not one that has been
observed in any experiments.  Note that the number of sub-bands used in this
figure (and most of the other figures presented) is substantially fewer than
would normally be used.  Generally the number of sub-bands required is about
20--50 depending on the exact parameters, however these cannot be well
distinguished from each other in a graph.  Therefore, throughout this paper,
most of the depictions of the growth of sub-bands will show only a few of
them.  

\begin{figure}
\begin{center}
\epsfig{file=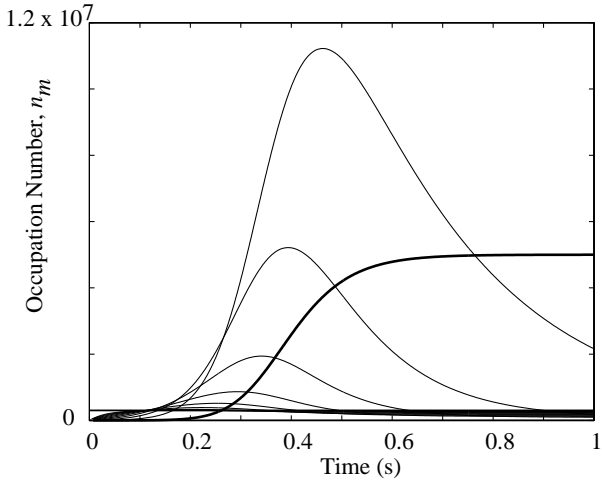,width=8cm}
\end{center}
\caption{Typical results of the new growth equations for the
growth of a Bose-Einstein Condensate using the same parameters as in
Fig.\ref{fig:simplegrowth}.  The condensate level (bold) reaches the
equilibrium population of $5 \times 10^6$ atoms, the other lines represent the
evolution of the populations of the other sub-bands in the condensate band. 
Note the much faster growth than in Fig.\ref{fig:simplegrowth}.}
\Label{fig:samplenewWplus}
\end{figure}

Once scattering processes are included (see below), reducing the number of
sub-bands used causes the growth to become slower.  It also causes the model to
become less realistic, since an individual level may then be described by an
average energy quite different from its actual energy.  If the number of
sub-bands is increased, an asymptotic limit to the speed of the growth is
reached, however this is also unrealistic since now some sub-bands contain only
fractions of individual energy levels.  The choice used in practice is such
that in their final equilibrium states the sub-bands have widths of
$\hbar\omega$.  This choice is close to the asymptotic limit, and ensures at
least three individual levels are contained in the first sub-band.

\subsubsection{Inclusion of Scattering Processes}
If we include scattering processes, as given by (\ref{scatt}), the picture is 
dramatically changed. Solutions for the resulting evolution
equations (\ref{totalgrowth}) are shown in
Fig.\ref{fig:sampletotal}.  This figure shows that the scattering has two
main effects.  Firstly, the initiation of the condensate level growth occurs
much more sharply, this gives a substantial change to the shape of the growth,
which has now lost much of the S-shape nature that previous solutions had
shown.  The speed of the growth after the initiation is changed little by the
inclusion of the scattering processes, since in this region growth is
completely dominated by the growth processes.  The second effect is that the
populations of the excited states no longer exceed plausible levels.

\begin{figure}
\epsfig{file=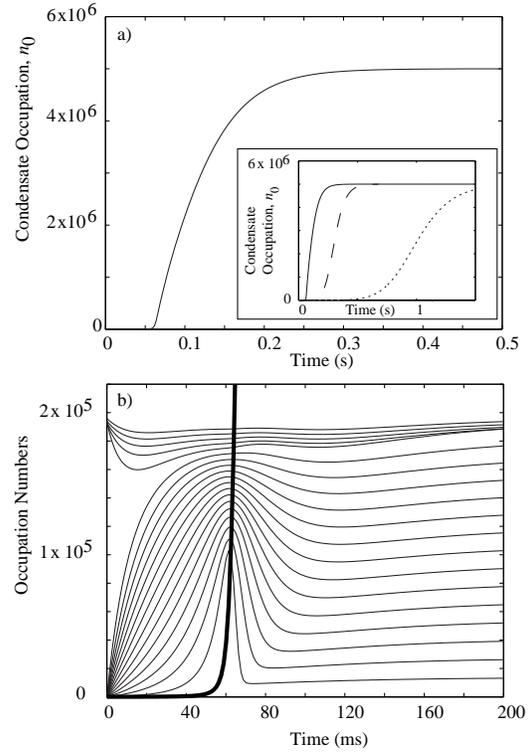,width=8cm}

\caption{Typical results of the total growth equation (including 
scattering processes) for the growth of a Bose-Einstein condensate 
using the same parameters as in Fig.\ref{fig:simplegrowth}.  a) The 
growth of the condensate (ground state) occupation.  The inset shows 
the same growth predicted by the total new model (solid) compared to 
the growth predicted without considering scattering processes 
(dashed), and that of the simple growth equation with an initial 
population of 100 atoms (dotted).  b) The growth of the explicitly 
considered sub-bands.  The condensate level itself is the bold curve.  
The top five sub-bands are those in the non-condensate band which 
were explicitly considered.  Note the different timescales on the 
graphs.}
\Label{fig:sampletotal}
\end{figure}

The reasons behind these changes are interlinked.  Without
scattering, all of the levels in the condensate band start to grow, and at
quite similar rates.  The difference in the growth rate between the very lower
energy states and that of the condensate level is particularly small. 
Because of the degeneracy of states in the lower sub-bands a large population 
can
form in them, which can become very much larger than the condensate level
population.  Once any one sub-band acquires a sufficient population the
stimulated term in the growth process begins to dominate and the population
increases even further.  In the absence of scattering processes, the only way
in which the excess population in these states can be transferred to the
condensate level, where it will be found in the equilibrium situation, is by a
transfer back to the non-condensate band followed by another collision which
transfers it directly to the condensate level.

If scattering processes are considered, atoms may now be transferred directly
between different levels in the condensate band in a collision.  Any excess
population in the excited states can then be quickly transferred out of the
state before the stimulated growth process becomes too dominant.

With the inclusion of the scattering processes, the effects of two important
approximations must be considered.  In the derivation of the equations
governing the scattering processes it was assumed that the value of $\Gamma(T)
\equiv \sum_{e_m>\Emax}e^{-e_m/kT}$ was equivalent to that for an
isotropic harmonic oscillator $\bar{\Gamma}(T)$, which affects the scattering
rates.  

The second important approximation was made in the solution of the equations
governing the growth processes, where the $W^{++}_m(n_0)$ terms were assumed to
be equal to $W^+(n_0)$ which has a value given by equation~(\ref{wplus}).  This
approximation has no effect on the growth rate of the condensate level if the
scattering terms are not considered, since then there are no interactions
between different levels in the condensate band. The effects of these two
approximations are discussed below.

\begin{figure}
\epsfig{file=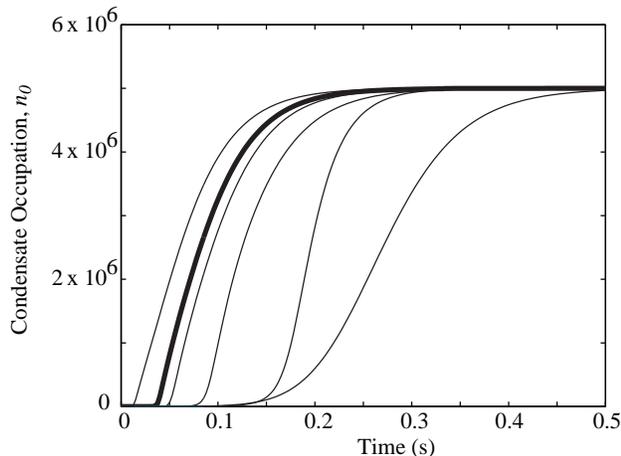,width=8cm}

\caption{The dependence on the rate for scattering processes of the 
condensate level growth.  Growth curves from left to right had 
factors $\Gamma(T)$ of $100\bar{\Gamma}(T)$, $\bar{\Gamma}(T)$, 
$\bar{\Gamma}(T)/2$, $\bar{\Gamma}(T)/10$, $\bar{\Gamma}(T)/100$ and 
zero respectively, where $\bar{\Gamma}(T)$ is the factor for the 
isotropic harmonic potential given by equation~(\ref{Gamma}).  These 
results were obtained using the same parameters as used in 
Fig.\ref{fig:sampletotal}.}
\Label{fig:modscattrate}
\end{figure}

\subsubsection{Effect of the Scattering Rate Approximation}

The magnitude of the scattering rate was assumed to be equal to that for an
isotropic harmonic oscillator in the derivation of equation~(\ref{scatt}).  In
Fig.\ref{fig:modscattrate} the growth of the condensate level is shown for
several different scattering rates.  The growth slows slightly if the rate is
decreased.   If the rate is increased, the growth becomes faster until it
reaches an asymptotic limit at which point it is the rate of the growth
processes which determines the speed of growth.  The results show that the
overall growth changes by only a relatively small amount (and certainly smaller
than present experimental uncertainties in growth experiments) provided that
the rate is within two orders of magnitude of that for the isotropic trap. 
Since it seems unlikely that the corrections due to the anisotropy would change
the rate by much more than one order of magnitude, the approximation of using
the isotropic rate factor seems to be valid.  It is interesting to note that a
scattering rate of only about 1\% of the isotropic case is usually sufficient
to prevent the occupations of the sub-bands becoming very large as they do in
the absence of any scattering.

\begin{figure}
\epsfig{file=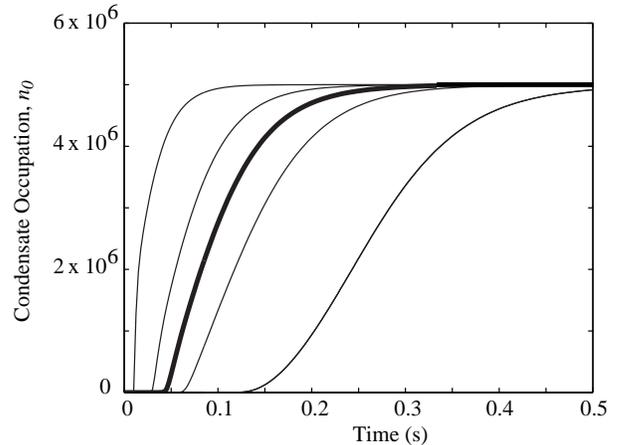,width=8cm}

\caption{The effect on the rate of growth of the condensate when the 
values of $W_m^{++}(n_0)$ were changed.  From left to right the 
curves correspond to values for $W_m^{++}(n_0)$ of $10W^+(n_0)$, 
$2W^+(n_0)$, $W^+(n_0)$, $W^+(n_0)/2$ and $W^+(n_0)/10$.  Conditions 
correspond to those in Fig.\ref{fig:sampletotal}.}
\Label{fig:modwplusrate}
\end{figure}

\subsubsection{Effect of $W_m^{++}$ factors}

The effect of the approximation $W_m^{++}(n_0) \approx W^+(n_0)$ is shown in
Fig.\ref{fig:modwplusrate} which shows the behaviour of the growth for
different values of the $W_m^{++}(n_0)$ terms.  Changing $W_m^{++}(n_0)$
changes the rate of growth of the excited states, and thus changes the
probability of atoms being scattered from the excited states into the
condensate level, giving a corresponding change in the overall growth of the
condensate. The results show that provided $W_m^{++}(n_0)$ lies in the range
$2W^+(n_0) > W_m^{++}(n_0) >W^+(n_0)/2$ then the growth rate does not change
significantly (compared to experimental uncertainties and the change caused by
using an accurate scattering rate).  However, outside of this range the growth
is altered considerably.  The expectation is that $W_m^{++}(n_0)$ will lie in
the desired range, since it is an average over quantities similar to
$W^+(n_0)$, and as such should be of the same magnitude.

The `standard' approximations for the rate constants that will be used in the
following results will be $W^{++}_m(n_0) = W^+(n_0)$ and $\Gamma(T) =
\bar{\Gamma}(T)$.

\subsubsection{Initial Conditions}
\Label{sec:initconds}

A problem which has to be considered in solving the total rate
equations~(\ref{totalgrowth}) is the determination of the correct initial
conditions.  Because the non-condensate band is assumed to be a thermal bath of
atoms at equilibrium, the initial populations for the explicitly considered
levels in this band are found from $[\exp((E-\mu)/kT) -1]^{-1}$.  Obviously
this cannot be used to give the initial populations in the condensate band when
the model is attempting to describe growth of the condensate.  It is not
immediately obvious what the appropriate initial conditions should be for the
condensate band.  In Fig.\ref{fig:initconds} four different initial
conditions are shown
\Item[a)] no initial population in the condensate band.  This is the most
artificial of the four possibilities presented.
\Item[b)] initial population of zero in the condensate level and excited state
occupations given by a linear dependence on energy rising to match that of the
non-condensate band at $\Emax$.
\Item[c)] initial populations of condensate band states all equal to the value
of $[\exp((\Emax-\mu)/kT) -1]^{-1}$.
\Item[d)] initial populations given by a linear extrapolation of the
non-condensate populations, meeting $[\exp((E-\mu)/kT) -1]^{-1}$ at 
$E_{R}$ tangentially. 
\Item[]In Fig.\ref{fig:initcondsgrowth} the growth curves corresponding to each of
the differing initial conditions are shown.  The different initial
conditions can be seen to have little effect on the shape of the growth, the
main effect is really just a small shift in the initiation time.  This effect
is generally quite small compared to the
effects of changing $\Gamma(T)$ and $W_m^{++}(n_0)$ as seen in the preceding
sections.  The fact that the initial conditions can be changed by so much and
yet have little effect on the growth curve is due mainly to the inclusion of
the scattering terms.  The scattering terms very quickly cause the levels in
the condensate band to come to a kind of quasi-equilibrium from whatever
initial state they are put in. 
Thus we conclude that an exact knowledge of the intial conditions is not 
important.  

Since the exact initial conditions do not seem to be important, in all further
calculations initial condition (c) will be used, as it is about mid-way between
the extremes of cases (a) and (d).
\begin{figure}
\epsfig{file=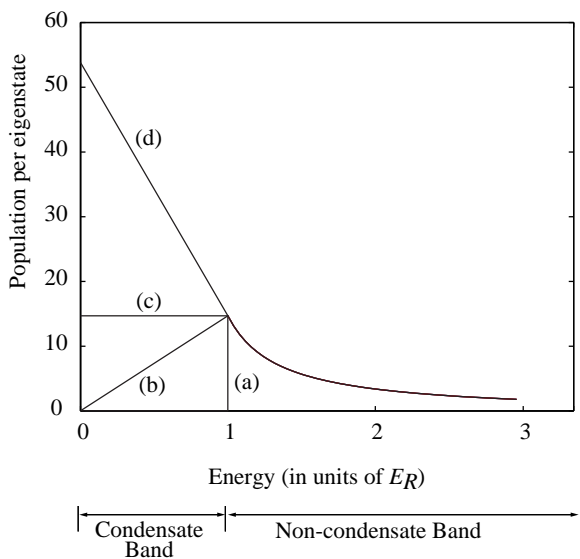,width=8cm}

\caption{Depiction of the four different initial populations trialed, 
as described in the text.}
\Label{fig:initconds}
\end{figure} 

\begin{figure}
\epsfig{file=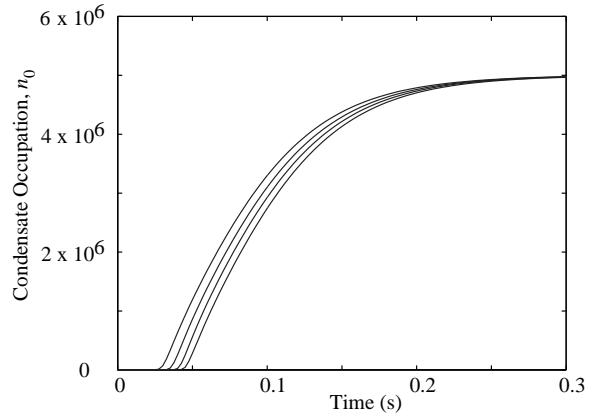,width=8cm}

\caption{Growth curves determined using the initial conditions 
described in the text.  The curves, from slowest to fastest growth, 
were given by initial conditions a-d respectively.  Conditions are 
the same as for Fig.\ref{fig:sampletotal}.}
\Label{fig:initcondsgrowth}
\end{figure} 

\subsubsection{Modifications of the Energies of the Sub-bands}
\Label{sec:modenergies}

The previous results were all obtained by taking the mean field effects due to
the condensate into account in the manner described in the previous section. 
That is to evenly distribute the energies of the sub-bands between the fixed
upper limit of $\ER$ and the $n_0$-dependent lower limit $\mu_C(n_0)$.  
Thus
all levels below $\ER$ (fixed at $2\mu_C(n_{0,f})$ where $n_{0,f}$ is
the final occupation of the condensate level) are modified at all times.  This
artificial model can be improved on, since the number of levels
affected by the condensate depends upon its occupation.  In other words, when
there is only a small condensate present it has a significant effect on only
the lower sub-bands, while the upper sub-bands are essentially unmodified.  An
improved (although still a little artificial) method is the following: all
levels below $\gamma \mu_C(n_0)$ (where $\gamma$ is an arbitrary parameter)
will be compressed to fit between this upper limit and the lower limit of
$\mu_C(n_0)$; the energies of all levels higher than $\gamma \mu_C(n_0)$ will
not be altered.  This is similar to the previous model, with the alteration
that the upper limit of the levels whose energies are changed is no longer
fixed, but instead rises with increasing $n_0$.  The energy of a level $e_m$ is
then given by
\begin{eqnarray}
e_m&=&\left\{ \begin{array}{ll}
e_m^0 & \mbox{for} \ e_m^0 > \gamma \mu_C(n_0) \\
\mu_C(n_0) + e_m^0 \left( 1 - {1\over \gamma} \right) & \mbox{for} \ e_m^0 <
\gamma \mu_C(n_0)
\end{array} \right., \Label{emdep}
\end{eqnarray}
where $e_m^0$ are the unchanged energy levels, as given by the eigenstates for
a non-interacting gas in a harmonic potential.  The energy spectrum of the
sub-bands given by this new model is shown in
Fig.\ref{fig:modenergyspectrum}, and compared to the previous energy
spectrum.  In the older model used so far in this paper all energy levels in
the condensate band are changed by the growth of the condensate.  However, the
extent of the condensate band is determined by the final occupation number of
the condensate (i.e., by the final value of $\mu_C(n_0)$).  In the new model, 
the
levels are modified in a more consistent fashion, the energy levels for any
given condensate population are determined by the condensate population at that
time, which appears to be a more logical approach. 

In section~\ref{sec:Emax} it was shown that a fair estimate for $\Emax$
was a value of $2\mu_C(n_{0,f})$.  The value for $\gamma$ will normally be
taken to be $2$, so that at equilibrium all energy levels below $\Emax$
will be modified.

\begin{figure}
\epsfig{file=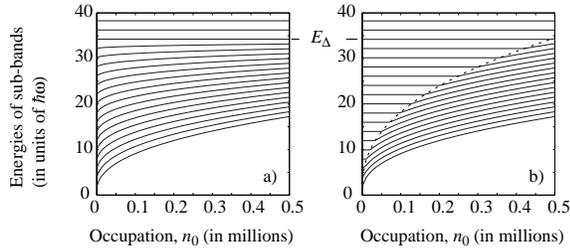,width=8cm}

\caption{The energy spectrum of the sub-bands as a function of $n_0$, 
modified to account for the mean field interactions with the growing 
condensate.  In (a) the levels are modified by changing all levels in 
the condensate band, as has been used to obtain the previous results.  
In (b) these effects are accounted for by the new model, explained in 
the text.  The lowest energy level is the condensate level, whose 
energy is $\mu_C(n_0)$.  The dotted line represents $2\mu_C(n_0)$, 
the maximum energy at which mean field corrections are assumed to be 
significant.  This figure shows the evolution of the energies for a 
temperature of $900{\rm nK}$ and with a final (equilibrium) 
condensate occupation of $5 \times 10^5$.  The few levels whose 
energies never change are those levels in the non-condensate band 
(above $\Emax$) which are explicitly considered.}
\Label{fig:modenergyspectrum}
\end{figure} 

The effect of this new model, compared to the earlier approach, on the overall
growth is very small, smaller in fact than any of the other effects discussed
in this section, and it is barely discernible.  This would seem to indicate
that it is the mean-field effects on the energies of the lower few sub-bands
which are important when considering the growth of the condensate, since in
both the old and new models the lower levels experience quite similar changes. 
Since the mean-field effects of the condensate on the higher levels seems to
have only a very small effect, the precise value of $\Emax$ would appear
to have little effect so long as it is reasonably high.  The value of
$2\mu_C(n_{0,f})$ seems to be a good value, since it is high enough that the
energy perturbations of higher energies do not have a large effect on the
growth, and it is low enough that the majority of atoms are found at higher
energies, giving an undepleted thermal bath.

\subsection{Comparison of Results with Experiment}
At the present point in time there has only been one published experiment which
has investigated in any detail the growth of a Bose-Einstein condensate.  The
experiment was performed at MIT using a trapped gas of $^{23}{\rm Na}$ atoms,
and the results were published in 1998 in ref~\cite{MITgrowth}.  

\subsubsection{MIT Experimental Method}
The MIT experiments were performed in the following way:  The gas of atoms was
confined in an approximately harmonic magnetic trap.  It was then cooled using
laser cooling and evaporative cooling techniques to a temperature slightly
higher than the critical temperature necessary for the formation of BEC.  At
this point the system is essentially in thermodynamic equilibrium, and then it
is suddenly put into a non-equilibrium configuration of lower energy, by means
of a rapid evaporative cooling `cut' which removes all atoms in states above a
certain energy in a time of about $10{\rm ms}$.  The system is then left to
relax to equilibrium with no further cooling.  The cut will have brought the
temperature of the gas below the critical temperature and so, to reach
equilibrium, a condensate will form.  The formation of the condensate is
observed at several stages during the evolution by the means of phase-contrast
microscopy.  

This method attempts to achieve, probably as closely as is realistically 
possible, an almost thermalised bath in contact with a condensate 
band, as has been assumed in our theoretical treatment.  The cut which 
removes the higher energy atoms causes the wings of the energy 
distribution to be truncated.  Experience with solutions of the 
quantum Boltzmann equation and related methods 
\cite{KS1997,HWC1997,QK2} shows that the effect of this will firstly 
be felt by the higher energy atoms remaining.  The higher levels will 
therefore be expected to thermalise more quickly, with thermalisation 
gradually moving through to the lower energies.  Thus at some point 
after the cut the majority of the atoms will be approximately in 
equilibrium, with the lower energy atoms still in very non-equilibrium 
states.

\subsubsection{MIT Experimental Results and Theoretical Comparisons}

Phase-contrast microscopy produces two dimensional images of the system with an
intensity proportional to the column density of the system.  From data of this
type temperatures, total numbers of atoms, and condensate level occupations
were extracted by the MIT group and were presented in~\cite{MITgrowth}.

The results found in~\cite{MITgrowth} were that condensate growth 
took on the order of $100$-$200{\rm ms}$ depending on the exact 
conditions.  It was found that the growth could be well fitted by a 
solution of the simple growth equation (\ref{simplegrowth2}), which 
can be put in the approximate (but in practice very accurate) form
\begin{equation}
\dot{n}_0 = \kappa_1 n_0\big[1-(n_0/n_{0,f})^{2/5}\big], \Label{bosstim}
\end{equation}
where, again, $n_{0,f}$ is the equilibrium condensate population.  The
solutions to this equation exhibit the S-shaped growth profile of 
Fig.\ref{fig:simplegrowth}.

The conclusion drawn
in~\cite{MITgrowth} was that a curve of this shape was evidence for the
importance of bosonic stimulation in the growth processes, since a purely
relaxational process would be described by solutions of
\begin{equation}
\dot{n}_0 = \kappa_2 (n_{0,f} - n_0). \Label{relax}
\end{equation}
However, the rate constants $\kappa_1$ found by the MIT group by 
fitting to the data obtained did not agree to better than an order of 
magnitude with the predictions of the simple growth equation of
\begin{equation}
\kappa_1 = 2W^+(n_0){\mu \over kT}.
\end{equation}
The simple growth equation did well to give the correct order of 
magnitude rates, however most of the predicted rates for higher 
temperatures seemed to be too small by about factor of three.  As the 
temperature decreased the discrepancy increased, the predicted rates 
became slower, whilst the experimentally fitted rate constants became 
larger.

\begin{figure}
\epsfig{file=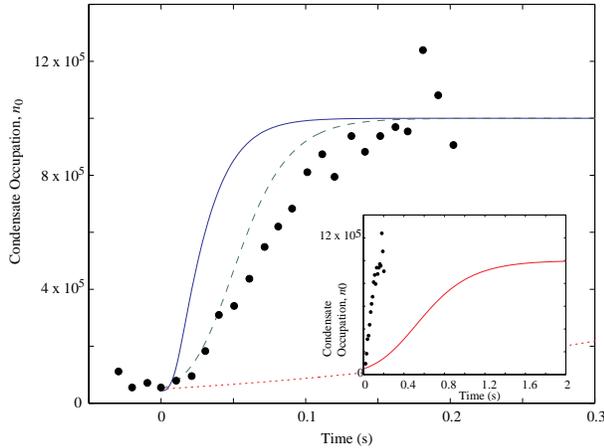,width=8cm}

\caption{Comparison of theoretical growth curves with data 
experimentally obtained at MIT (dots).  The measured temperature for 
the experimental data was $1200{\rm nK}$.  In the main part of the 
figure the solid line shows the theoretical fits with standard rate 
constant approximations at $1200{\rm nK}$, the dashed line shows the 
theoretical curve obtained by using $\bar{\Gamma}(T) = \Gamma(T)/10$ 
and $W_m^{++}(n_0) = W^+(n_0)/2$.  The dotted line shows the growth 
predicted by the simple growth equation of~\protect\cite{BosGro} 
(equation~(\ref{simplegrowth})), which is shown on a larger scale in 
the inset.  The initial condensate populations for each curve were 
set to $5 \times 10^4$, the experimental value at $t=0$.  The origin 
in the time axis represents the time at which the quick cooling `cut' 
in the experiment was finished.}
\Label{fig:mitcompare1200nK}
\end{figure}

In contrast, the solutions to the growth model presented in this paper no
longer show the S-shape, but are in fact closer in shape to the
solutions of equation~(\ref{relax}).  These curves were found by MIT to
describe the data quite poorly if the growth started at time $t=0$ (the time at
which the cut finished).  However, if an initiation time was allowed before the
growth began, such solutions became quite close fits, although they still did
not describe the initially slow growth giving the S-shape to the growth.  The
results obtained in this paper show that, while the general shape is that of
solutions to equation~(\ref{relax}), there is also an initiation time present
before the growth begins.

As a specific comparison with experiment, 
Fig.\ref{fig:mitcompare1200nK} shows the comparison between the growth 
curve predicted here and the experimental data.  The 
data is similar to that in~\cite{MITgrowth}, but was 
not actually published.  It was provided by MIT as being the growth of 
a condensate with an equilibrium population of about $1 \times 10^6$ 
atoms, at a temperature of $1200{\rm nK}$~\cite{MITpersonal}.

In the growth experiments statistical uncertainties are 
estimated~\cite{Wolfgang} to be 10\% 
for relative number measurements, 15\%
for temperatures.  Systematic uncertainties are estimated
as 20\% 
in absolute number measurements, and 8\%
for temperature measurements.
Condensate occupations of less than 
$10^5$ atoms could not be discerned against the background of the 
thermal vapour cloud~\cite{MITgrowth}.

As Fig.\ref{fig:mitcompare1200nK} shows, the growth predicted by our model
is quite a good fit to the data, and the order of magnitude is certainly
predicted well.  This is a substantial improvement to the growth predicted by
the simple growth equation at the same conditions
which gives growth over about $1.5$ seconds, as opposed to the experimental
results of about $0.15{\rm s}$.  At the stated parameters the theoretical fit
could still be improved.  The dashed line in Fig.\ref{fig:mitcompare1200nK}
indicates that by adjusting the scattering rate and values of $W_m^{++}$ (as
was discussed in previous sections) a better fit may be obtained.

\begin{figure}
\epsfig{file=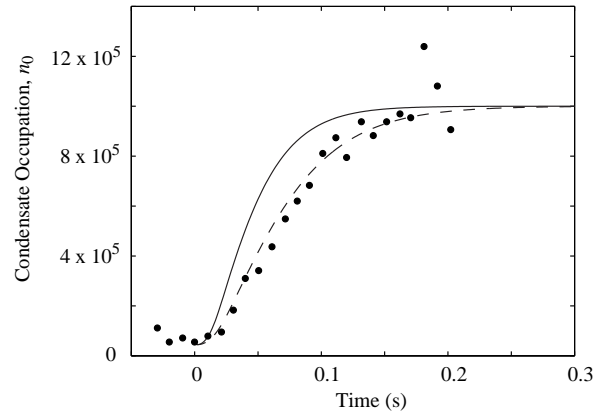,width=8cm}

\caption{Comparison of theoretical growth with MIT experimental data.  The
experimental data is the same as that used in
Fig.\ref{fig:mitcompare1200nK}. The theoretical curves were determined by
using temperatures of $1000{\rm nK}$ (solid) and $850{\rm nK}$ (dashed).  The
rate constants were taken to be their standard values of $\bar{\Gamma}(T) =
\Gamma(T)$ and $W_m^{++}(n_0) = W^+(n_0)$.  Again, the initial condensate
populations for each curve were set to $5 \times 10^4$.}
\Label{fig:mitcompareother}
\end{figure}
The temperature plays a sensitive role in these comparisons, since the 
rate of growth is quite sensitive to temperature.
%
%
%
Furthermore, as will be seen in the following section, the fitting 
method may play a significant role.  In Fig.\ref{fig:mitcompareother} 
the theoretical curve is plotted (using the standard values of the 
scattering rate and $W_m^{++}$) for two lower temperatures.  It can be 
seen that the fit is very good for the $850{\rm nK}$ results, and not 
quite so good, but still quite close, in the case of the $1000{\rm 
nK}$ curve.

This highlights a difficulty in comparing theoretical predictions with 
experimental measurements of condensate growth.  The spatial density 
distribution of the thermal cloud (from which the temperature is 
measured) changes only slightly with temperature, whereas the growth 
rate is quite strongly dependent on the temperature.  There are other 
problems as well, the most prominent of these being the difficulty of 
experimentally determining the condensate occupation.  The spatial 
distributions of particles in the first few excited states are quite 
narrow functions, and they can overlap the condensate level 
distribution significantly.  It then becomes extremely difficult to 
distinguish condensate level atoms from low energy excited state atoms 
in experimental measurement.  If the measured `condensate occupation' 
was in fact the occupation of the lowest 5 levels (for example) this 
would alter the growth curve in some important regions.  The main 
effect would be that the growth would appear more gradual during the 
early times, and would not seem to have such a sharp initiation.  This 
is a possible explanation for the S-shape that was found in the MIT 
data.  Clearly what is needed are theoretical predictions for the 
overall spatial density distributions during the growth, rather than 
merely occupation numbers for the various states.  This will be 
addressed in the following section.

\begin{figure}
\epsfig{file=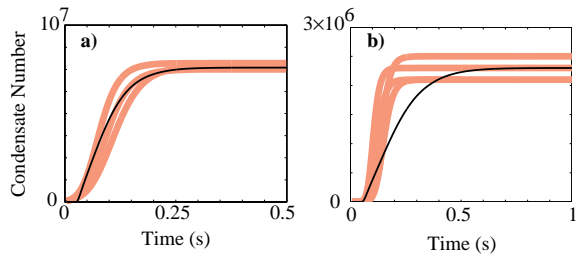,width=8cm}

\caption{Comparison of theoretical growth (thin lines) with curves
fitted by MIT to experimental data in~\protect\cite{MITgrowth} (broad lines). 
The MIT curve fits are represented as broad lines to indicate that they are
fitted curves with unknown (but probably substantial) uncertainties. {\bf
a)}~Theory $T=830{\rm nK}$, $n_{0,f}=7.6\times10^{6}$;  Experiment 
$T=810-890{\rm nK}$, $n_{0,f}=7.5-7.85\times10^{6}$; {\bf b)}~Theory $T=590{\rm
nK}$, $n_{0,f}=2.3\times10^{6}$;  Experiment  $T=580-610{\rm nK}$,
$n_{0,f}=2.1-2.5\times10^{6}$.  The initial populations were treated as free
parameters to best match the initiation of the growth with the MIT curves.}
\Label{fig:mitgrowthcompare}
\end{figure} 

The majority of the data contained in~\cite{MITgrowth} was presented in the
form of rate constants for fits  to experimental data of the type described by
equation~(\ref{bosstim}).  Fig.\ref{fig:mitgrowthcompare} shows the
comparison of our theoretical growth curves as compared to solutions of
equation~(\ref{bosstim}) using a selection of the MIT fitted constants
$\kappa_1$.  The figure shows good overall agreement with the experimental
data.  The agreement is better at higher temperatures.  At the lower
temperatures reached in the experiment the agreement is less good.

However, the results of this paper still show an overall decrease in 
the the rate of growth with decreasing temperature.  This is the 
opposite trend to that experimentally observed.  This could partly be 
due to uncertainties in extracting numbers from the experimental 
data.  Another possible explanantion is that in order to cool to such 
low temperatures from just above the critical temperature in 10ms a 
large proportion of the atoms must be removed.  This will give rise to 
a system in a highly non-equilibrium state, the relaxation from which 
may be inadequately described by our model.

%
%
\section{Spatial Density Distribution of a Condensate 
System}\label{spatial}

In the MIT experiments into condensate growth~\cite{MITgrowth} the 
data is collected by phase-contrast microscopy 
\cite{MITphasecontrast1,MITphasecontrast2}, the measurement of change 
in the phase of light after it has passed through the vapour cloud.  
The result is a two dimensional plot of the column density integrated 
along the third dimension.  As discussed in the previous section, the 
extraction of the population per energy level from this experimental 
data is complicated.  Therefore, in order to more easily enable 
comparison with experiment, it is desirable to obtain from our 
theoretical results predictions of the spatial distribution of the 
condensate as it grows.  When the system is in equilibrium this 
distribution is well known, but this is not the case during the 
condensate growth that we are interested in.

\subsection{Semiclassical Phase-space Description}

In order to convert the results of the model into spatial distributions, the
spatial probability distributions for each energy band need to be found.  The
distribution for the ground state $\rho_{\rm gs}(\br)$ is well described by
the Thomas-Fermi approximation for the wave function:
\begin{eqnarray} \Label{eq:TFgs}
\rho_{\rm gs}(\br) &=& \left| \psi_{TF}(\br) \right|^2
\nonumber \\
& =& \frac{m}{4\pi\hbar^2a}
\left[ \mu_C(n_0) - V_T(\br)
\right] \theta \left( \mu_C(n_0) - V_T(\br) \right),
\end{eqnarray}
where $V_T(\br)$ is the trapping potential, and $\theta(x)$ is the 
step function.  This is a very good approximation to the shape of the 
condensate when $n_0$ is large, failing only at the very edge of the 
condensate where the numerical solution to the Gross-Pitaevskii 
equation vanishes smoothly.

A description of the spatial distributions for each of the higher energy
sub-bands can be found by using a semiclassical phase-space approach.  The
cumulative number of states below an energy $E$ is given by
\begin{eqnarray}
{\rm N}(E) ={1\over h^3}\int d^3\br\,\int d^3\bP\,\, \theta \left( 
E - {P^2\over2m} -
V_{\rm NC}^{{\rm eff}}(\br) \right) ,
\end{eqnarray}
where $V_{\rm NC}^{{\rm eff}}(\br)$ is the
effective potential experienced by an atom.  This gives a density of states
\begin{eqnarray} \Label{eq:densstates}
g(E) &=& {d{\rm N}(E)\over dE}
\nonumber \\
& = &{1\over h^3}\int d^3\br\,\, \int d^3\bP\,\, 
\delta \left( E -
{P^2\over2m} - V_{\rm NC}^{{\rm eff}}(\br) \right). 
\end{eqnarray}
Using the local density approximation, for an energy sub-band with energy $e_m$
and width $\Delta e_m$, the average spatial distribution atoms in the band
(averaged over all the wavefunctions of all the states in the band) may be
obtained by removing the $d^3\br$ integral and integrating over $d^3\bP$.  This
gives
\begin{eqnarray} \Label{eq:rhoEofr}
\rho_m(\br) &=& {1\over h^3}\int d^3\bP\, \delta \left( e_m -
{P^2\over 2m} - V_{\rm NC}^{{\rm eff}}(\br)\right) \, \Delta e_m 
\\
\Label{eq:spatdist}
 &=& \left(4 \pi2^{1/2}m^{3/2}/h^3 \right)\,
 \Delta e_m \, \sqrt{e_m - V_{\rm NC}^{{\rm eff}}
(\br)}. 
\end{eqnarray}

\begin{figure} 
\begin{center}
\epsfig{file=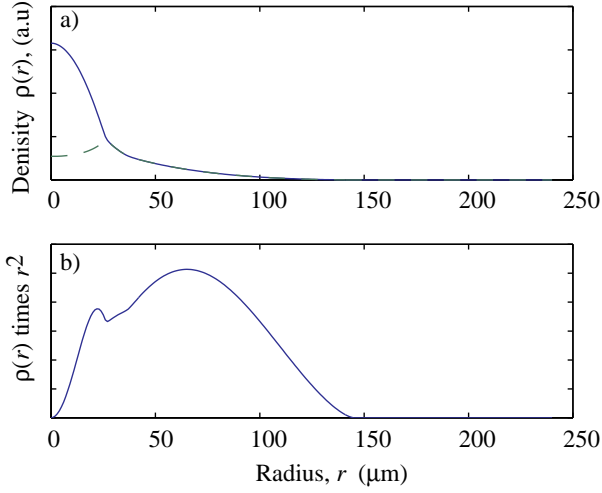, width=8cm}
\end{center}

\caption{Sample radial spatial density distribution for condensate 
having reached thermal equilibrium.  a) Solid---total density 
distribution, dashed---density distribution due to excited states 
only.  b) Density times radius squared.  These results are for the 
trap parameters used in the MIT growth experiments 
\protect\cite{MITgrowth,MITpersonal} }\Label{fig:sampleshape}
\end{figure} 

\subsubsection{Resulting Spatial Density Distribution}
We will use the semi-classical distributions for the excited states, and the
Thomas-Fermi wavefunction for the condensate. The total radial probability
distribution may be calculated by normalising each level, and then summing over
all the levels.  The distributions are normalised so as to give the appropriate
population in each level.  For the bands below $\Emax$ these populations
are given by the numerical results of our model.  For those
above $\Emax$ the population is determined by the Bose-Einstein
distribution function $F(E) = [\exp((E-\mu)/kT) - 1]^{-1}$.  This gives the
spatial density distribution for the whole condensate system in three
dimensions.

An example of the resulting spatial density distribution for a system at
equilibrium is shown as Fig.\ref{fig:sampleshape}.  In the first part of
this figure the total density is plotted as a function of radius, as well as
the density due to the excited states only.  The density due to the excited
states can be seen to be significantly decreased in the region of the
condensate.  In Fig.\ref{fig:sampleshape}b the density multiplied by the
radius squared is plotted, this is proportional to the total number of atoms
found at any radius (due to the three dimensional nature of the distribution). 
From this it can be seen that, even though the density in the centre due to the
condensate is much larger than elsewhere, the majority of the atoms are still
found in the surrounding vapour cloud (as was assumed in the derivation of the
model).

When the system is in equilibrium these results can be checked.  At equilibrium
the non-condensate spatial distribution can be obtained directly  from
equation~(\ref{eq:densstates}).  The total number of non-condensed atoms, $M$
is
\begin{eqnarray}
M &=& \int_0^{\infty}g(\varepsilon){1 \over e^{(\varepsilon -\mu)/{kT}}-1} 
d\varepsilon 
\end{eqnarray}
which corresponds to the local density form of $\rho(\br)$
\begin{eqnarray}  
\rho(\br)  &=& {1\over h^3}\int d\varepsilon\int d^3\bP\,\, 
{\delta \left( E - {P^2/2m} -
V_{\rm NC}^{{\rm eff}}(\br) \right)\over e^{(\varepsilon -\mu)/{kT}}-1}
\nonumber \\
\\
&=& {1\over h^3}\int d^3\bP\,\,\left\{ \exp \left[{{{P^2 \over 2m} +
V_{\rm NC}^{{\rm eff}}(\br)
-\mu \over {kT}}}\right]-1 \right\}^{-1} .
\nonumber\\ 
\end{eqnarray}
Carrying out the integral over momentum space gives
\begin{eqnarray} \Label{eq:g32}
\rho(\br) &=& \left( {mkT \over 2\pi\hbar^2} \right)^{3/2} 
G_{3/2}\left( {V_{\rm NC}^{{\rm eff}}(\br) - \mu \over {kT}}\right) \\
&&\ \ \ \ \mbox{where} \ G_{\sigma}(z) \equiv 
\sum_{n=1}^\infty n^{-\sigma} e^{-n z} 
\end{eqnarray}

A comparison of the non-condensate density obtained using 
equation~(\ref{eq:g32}) and that calculated by summing 
equation~(\ref{eq:spatdist}) over $1200$ levels is shown in 
Fig.\ref{fig:shapevsg32}.  It can be seen that the agreement 
between the two methods, for this \emph{equilibrium} situation, is 
good.  The agreement improves if more energy levels are included in 
the sum over equation~(\ref{eq:spatdist}), and using about $2000$ 
levels gives very good agreement.  Our semiclassical method therefore 
shows good agreement with the expected distribution at equilibrium, 
and this is the only case in which we can be certain of the 
theoretically correct result.

In obtaining the results in this section we have only considered 
effects of the mean-field repulsion due to the condensate atoms, both 
on the thermal vapour cloud and on the condensate itself.  In order to 
be truly consistent the model should also include the effects of the 
mean-field repulsion due to the thermal cloud on the system.  This is 
quite easily achieved mathematically, but it does increase the 
computational time by a very substantial amount.  However, recently 
Naraschewski and Stamper-Kurn~\cite{NSK1998} compared the density 
distributions obtained both with and without considering the 
mean-field repulsions of the thermal cloud, for an equilibrium 
condensate system.  They found that the overall density distributions 
for the two cases were practically indistinguishable except for a very 
small deviation at the edge of the condensate.  Furthermore, 
 Holzmann, Krauth and Naraschewski~\cite{HKN1998} found 
that the semiclassical density distribution (including the mean-field 
repulsions by the thermal cloud) gives excellent agreement with exact 
Quantum Monte Carlo simulations for dilute gas condensates in 
equilibrium.  Thus we expect that the density given by our 
semiclassical method is a very good description of the realistic 
system.

\begin{figure} 
\begin{center}
\epsfig{file=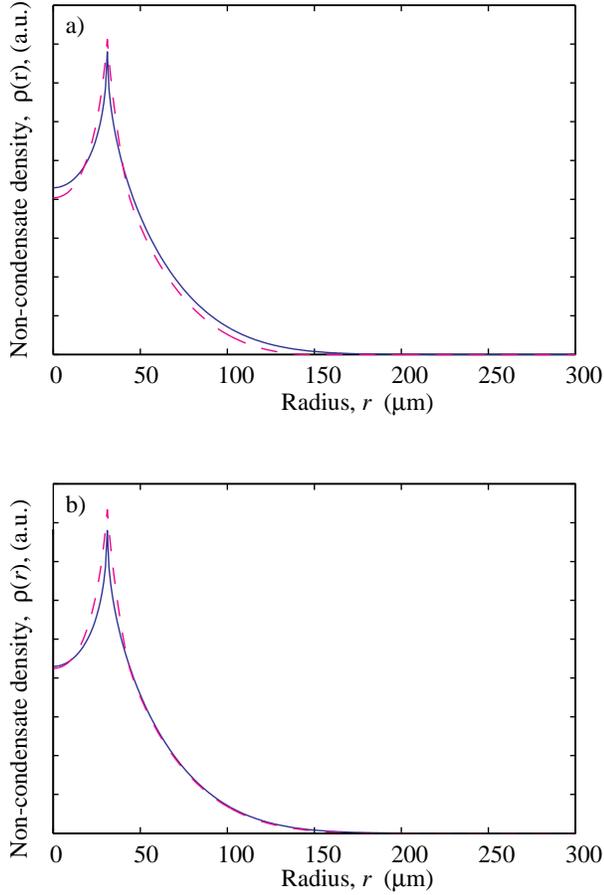,width=8cm}
\end{center}

\caption{Comparison of the non-condensate density
distribution at equilibrium calculated a) from the sum of $1200$ individual
energy level distributions (dashed), and using the $G_{3/2}(z)$ function
(solid); b) as for (a) but using the sum of $2000$ energy levels.}
\Label{fig:shapevsg32}
\end{figure}

\subsubsection{Description of a Realistic Experimental System}

The next step that must be performed, in order to compare with the experimental
results, is to perform a column integral along one dimension, yielding a
function of two spatial dimensions only.  This is relatively straightforward to
achieve numerically.

Finally the asymmetry of the real traps must be taken into account.  The
previous arguments for the exact non-interacting wavefunctions assumed the use
of a spherically symmetric trap, whereas the realistic traps used are strongly
anisotropic in the $z$ dimension.  However, in all the previous semiclassical 
arguments the
only effect is to change $V_T$ from $m\omega^2r^2/2$ to
$m\omega_{xy}^2(x^2+y^2)/2 + m\omega_z^2z^2/2$ where now
\begin{eqnarray}
\omega_{xy} &=& \sqrt{\lambda}\bar{\omega} \\
\omega_z &=& \bar{\omega}/\lambda \\
\bar{\omega} &=& (\omega_x\omega_y\omega_z)^{1/3} \\
\lambda &=& (\omega_{x}/\omega_z)^{2/3} .
\end{eqnarray}
Due to the harmonic nature of the trap, one can
recover the original form of the potential by scaling the dimensions used. 
Defining 
\begin{eqnarray}
\bar{z} &=& z/\lambda \\
\bar{x} &=& x\sqrt{\lambda} \\
\bar{y} &=& y\sqrt{\lambda},
\end{eqnarray}
the potential now returns to the form $V_T(\bar{\br}) =
m\bar{\omega}^2\bar{r}^2/2$, where the scaled radius is given by $\bar{r} =
\sqrt{\bar{x}^2 + \bar{y}^2 + \bar{z}^2}$.  Equation~(\ref{eq:spatdist}) can now
still be used, and the column density integration performed as for an isotropic
potential, and the resulting function of $\bar{x}$ and $\bar{z}$ needs only to
be rescaled to recover the answer in terms of $x$ and $z$.  This scaling also
affects the numerical column integration, with the effect that the result needs
to be scaled overall by a factor of $\lambda^{-3/2}$ (assuming the integration
is in one of the two shorter dimensions and that the result is in the
asymmetric (longer) dimension, as is the case for the `cigar' geometry traps of
MIT).

\subsection{MIT Fitting Method}

In the MIT experiments, the data for the numbers of condensate atoms 
and temperatures are obtained from fitting to the density profile 
obtained.  The raw data obtained is in the form of two dimensional 
images and, although these can be fitted, in order to save time and 
computational resources the fit was mostly performed only to a one 
dimensional slice through the center of the 
condensate~\cite{MITpersonal}.  This density profile is fitted from a 
function formed by the combination of a condensate density profile 
and non-condensate profile.  The MIT fitting procedure (presented in 
\cite{MITreview}) is a phenomenological procedure, only loosely 
connected with fundamental theory.  It is based on two observations:

\begin{itemize}

\item[i)]  The behavior of the thermal cloud in the wings of the 
profile is almost independent of the chemical potential---its behavior 
in the wings can thus be used to determine the temperature of the 
vapor.  

\item[ii)] The center of the profile is dominated by the condensate, 
and this---after subtracting the contribution of the vapor---can be 
fitted to the parabolic Thomas-Fermi profile, to determine the 
condensate chemical potential and thus the number in the condensate.
\end{itemize}

In order to execute the second procedure, some estimate of the vapor 
contribution has to be given in the region where the condensate is 
present.  This is done by fitting the profile to the sum of a 
Thomas-Fermi parabola and a zero chemical potential Bose-Einstein 
distribution.

Thus, using our methodology, this would be as follows.
Setting the chemical potential to zero and
integrating along one dimension gives (for a slice where $x=0$)
\begin{eqnarray}
\rho_{\rm int}(z,x=0) &=& \int_{-\infty}^{\infty} \rho(z,y,x=0) dy \\
&=& {m(kT)^2 \over 2\pi\hbar^3\omega_y} G_2 \left(  {m\omega_z^2 \over 2kT} 
z^2 \right). \Label{eq:MITnoncondfit}
\end{eqnarray}
This function can then be used to fit the wings of the distribution to obtain
$T$. 

The column density function due to the condensate, $\rho_{\rm int, 
C}$, is then obtained by integrating the Thomas-Fermi wavefunction 
given by equation~(\ref{eq:TFgs}) over one dimension
\begin{eqnarray}
&&\rho_{\rm int,C}(z,x=0) 
\nonumber \\
&&\qquad
= \int_{-y_1(z)}^{y_1(z)} 
\left[ {\mu_C(n_0) \over u}-{m \over 2u}( \omega_z^2 z^2 + \omega_y^2 y^2)\right] dy 
\nonumber \\  
\\
&&\qquad
\mbox{where} 
\ \ y_1(z) = \sqrt{{2 \mu_C(n_0) \over m\omega_y^2} - 
{\omega_z^2\over \omega_y^2}z^2}, 
\end{eqnarray}
and where $u = 4\pi\hbar^2 a/m$, giving
\begin{eqnarray} \Label{eq:MITcondfit}
\rho_{\rm int,C}(z,x=0) &=& {2 m \omega_y^2 \over 3u}
\left( {2 \mu_C(n_0) \over m\omega_y^2} - {\omega_z^2
\over \omega_y^2}z^2 \right)^{3/2}.
\end{eqnarray}
The density profile can therefore be fitted by the sum of
equation~(\ref{eq:MITcondfit}) and equation~(\ref{eq:MITnoncondfit}).

In the measurement procedure used, the phase shift data was not 
calibrated independently.  Instead, the raw data were fitted using 
semiphenomenological fit functions and an arbitrary scale factor to 
related the measured phase shif to the column density.  

However, the actual procedure used is somewhat more general, since 
the MIT procedure uses the trap constants as further fitting 
parameters, and these are allowed to be {\em independently} chosen 
for the condensate and the thermal cloud.  This is done since finite 
resolution effects of the imaging system may have a rather different 
effect on the very narrow condensate profile from that on the broader 
vapor profile.  In practice, for similar reasons the scale factor 
relating column density to the phase-contrast signal is allowed to be 
different for vapor and condensate.  The number of condensate atoms 
$n_0$ can be obtained from fitting directly to 
equation~(\ref{eq:MITcondfit}), or, by a simpler method used for most 
cases, determining the extent of this condensate distribution.  In 
most cases, since it is computationally simpler, number of condensate 
atoms was obtained from the spatial width of the condensate $z_0$.  
These are related by \begin{eqnarray}&& \rho_{\rm int,C}(z_0,x=0) = 0 
\ \ \mbox{when} \ \ z_0^2 = {2\mu_C(n_0) \over m\omega_z^2} 
\nonumber\\
&&\qquad= \left( {15 u (m/2)^{3/2}\bar{\omega}^3 \over 8 \pi}
\right)^{2/5} {2 n_0^{2/5}
\over m\omega_z^2} .
\Label{condextent}
\end{eqnarray}
thus $n_0 \propto z_0^5$.

In our fits we have assumed the scale factor to be the same for 
condensate and vapor, since our model is physically based, rather 
than being semiphenomenological.  Hence, in our procedure, there are 
three free parameters to be fitted: $T$, $n_0$ and a scale factor.

\subsection{Comparison of Fitting Methods}

In Fig.\ref{fig:MITfitting} we show a least squares fit of a zero chemical 
potential plus a Thomas-Fermi condensate profile for an equilibrium 
distribution.  The data used was obtained from Fig.2 
of~\cite{MITgrowth}, for which the MIT group obtained values of $T = 
800{\rm nK}$ with 9 million condensate atoms present in equilibrium 
after $160{\rm ms}$~\cite{MITpersonal}.  Also shown is the 
contribution to the distribution due to the non-condensate atoms only.  
Note that the $z$ axis has been scaled by a factor of $3.2$ over that 
in~\cite{MITgrowth} due to a typographical error in the 
paper~\cite{MITpersonal}.

In this fit, as for all the following cases, the density scale factor 
(necessary to convert from the arbitrary unit scale resulting from the 
experimental measurements) was determined by the least squares fitting 
procedure.  Since we use the $T$ and $n_0$ determined by the MIT 
group using their procedure, this scale factor is the only free 
parameter in this fit.  

\begin{figure} 
\epsfig{file = 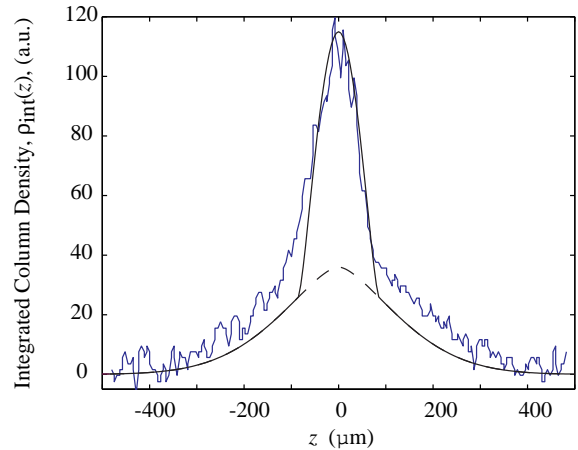,width=8cm} 

\caption{Equilibrium data obtained from MIT~\protect\cite{MITgrowth} 
fitted using the MIT procedure as described in the text (solid smooth 
curve) and showing the component due to the non-condensate (dashed).  
Parameters used were $T = 800 {\rm nK}$, 9 million condensate atoms, 
and a density scale factor of $1.76 \times 10^{-14}$.  
}\Label{fig:MITfitting}
\end{figure}

Because the non-condensate density is taken to be given by a 
distribution with zero chemical potential, which means in part that 
the mean field interactions with the condensate are neglected (as now 
$V_{\rm eff}(\br) = V_T(\br)$), this allows the non-condensate 
distribution to be quite strongly peaked at small $z$, whereas the 
inclusion of this effect (as shown in Fig.\ref{fig:sampleshape}) 
causes the distribution in this region to be suppressed.  This could 
lead to an underestimate of the number of condensate atoms, which 
might be particularly significant at the initial stages of the 
condensate formation due to the low numbers present.  Using the full 
MIT procedure with up to 5 extra parameters must give a better fit, 
but this basic problem will still be present, since the qualitative 
behavior of the vapor in the condensate region will be very similar to 
this fit.

\begin{figure} 
\epsfig{file=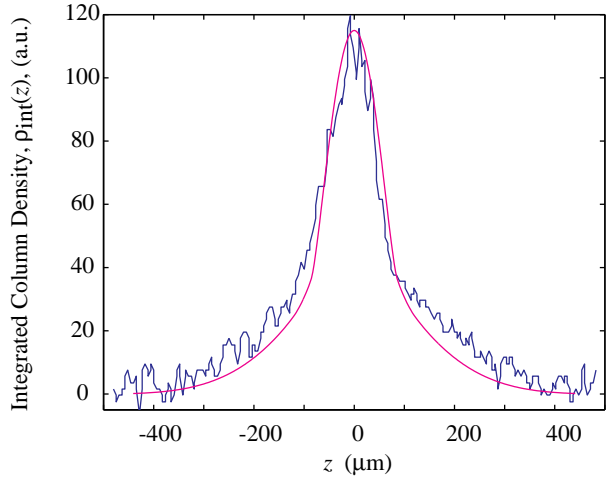,width=8cm}

\caption{Equilibrium data as for the previous figure, fitted by using 
the semiclassical method proposed in this paper.  Temperature and 
number of condensate atoms were taken to be the same as was used 
previously ($T = 800 {\rm nK}$ and $n_0 = 9 \times 10^6$), however 
the density scale factor was changed to $4.72 \times 10^{-14}$.}
\Label{fig:ourfittheirparam}
\end{figure}

The same data can be fitted using the distributions from 
equation~(\ref{eq:spatdist}).  Figure~\ref{fig:ourfittheirparam} shows 
the same equilibrium data, fitted by using the same parameters but by 
using the semiclassical distribution---as before, we use the $T$ and 
$n_0$ determined by the MIT group using their procedure, so that the 
only free parameter is the scale factor.  This spatial distribution 
can be seen to fit the experimental data much more closely.  However 
this can still be improved using a different choice of $T$ and $n_0$.  
The best fit was found with parameters of $T = 900 {\rm nK}$ with $4$ 
million condensate atoms and a density scale factor of $5.4 \times 
10^{-14}$, and is shown in Fig.\ref{fig:myfitmyparam}.  As a 
measure of how good the fit was, the sum of the square of the 
differences between the data and the fitted curve was calculated.  The 
best curve gave a value of $5.7 \times 10^3$, the fit in 
Fig.\ref{fig:ourfittheirparam} gave a value of $1.23 \times 10^4$ 
and the  fit of Fig.\ref{fig:MITfitting} gave $1.76 \times 10^4$.

\begin{figure} 
\epsfig{file = 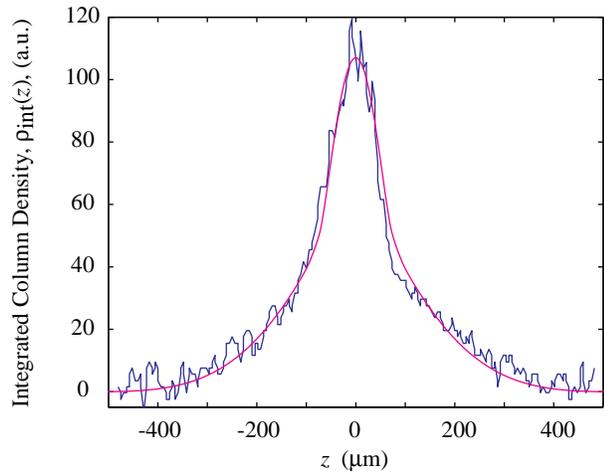,width = 8cm}

\caption{The best fit to the same data used in the previous two 
figures.  The parameters used were $T = 900 {\rm nK}$, $n_0 = 4 
\times 10^6$ and the density scale factor was $5.4 \times 10^{-14}$.}
\Label{fig:myfitmyparam}
\end{figure}

The temperature determined by our method at $900\rm K$ is not very different
from the $800\rm K$ determined by the MIT procedure, but the 
condensate number by our method is at $4\times 10^6$ less than half 
of the MIT value of $9\times 10^6$. Note however that the fit of 
Fig.\ref{fig:ourfittheirparam} does not appear visually 
unacceptable, leading to the conclusion that the best we can say is 
that the data do not constrain the condensate number very strongly. 
The reason for this probably lies in the choice of the least squares fitting 
procedure, which weights all data equally, and since there is not very 
much data in the condensate region, the condensate number can be 
varied significantly without changing the goodness of fit greatly.

In order to appreciate better the differences between the three fits, 
we have plotted all three fit functions on the same graph in 
Fig.\ref{fitcompares}. From this it is clear that the fits using the 
zero chemical potential Bose-Einstein Fig.\ref{fitcompares}(a)
and the more correct  Fig.\ref{fitcompares}(b) are much more similar 
to each other than to the best fit, Fig.\ref{fitcompares}(c), in which 
the higher temperature makes for a much more pronounced peak in the 
vapor distribution, upon which a condensate peak is then superposed, 
giving a larger overall peak than would be expected from the 
condensate.  From this figure, it is clear that the properties of the 
distribution in the transition region can be very deceptive.

We should also note that the density scale factor does not change 
significantly from Fig.\ref{fig:ourfittheirparam} to 
Fig.\ref{fig:myfitmyparam}.  Since this scale factor is a well 
defined physical quantity which should be determined more or less 
equally by all data, this is encouraging.  By this we mean that, 
although the scale factor is of no particular interest to the physics 
of Bose-Einstein condensation, it is nevertheless a well defined 
physical quantity whose value should be determinable.  In contrast the 
fit of Fig.\ref{fig:MITfitting} gives a very different scale 
factor, which is not unexpected, given that it does not pass through 
most of the data points.

\begin{figure} 

\epsfig{file = 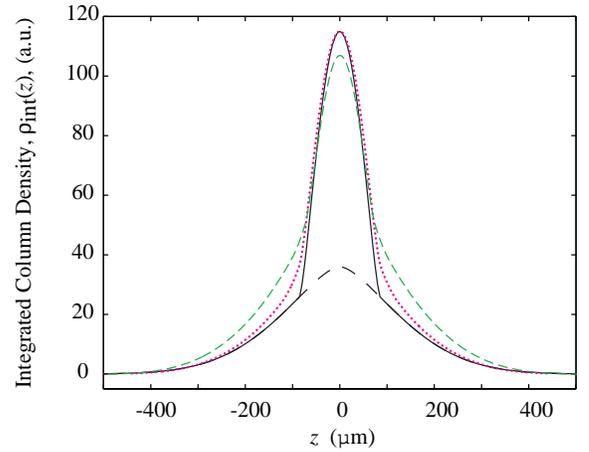,width = 8cm}
\caption{Comparison of the three fit functions: (a) Solid line, The 
zero chemical potential vapor fit as in Fig.\ref{fig:MITfitting}; 
(b) Dotted line, Fit using a nonzero 
chemical potential vapor as in Fig.\ref{fig:ourfittheirparam}; (c) 
Dashed line, The best fit as in Fig.\ref{fig:myfitmyparam}}
\Label{fitcompares}
\end{figure}

\EndTwoColumn

\begin{figure} 
\begin{center}
\epsfig{file = 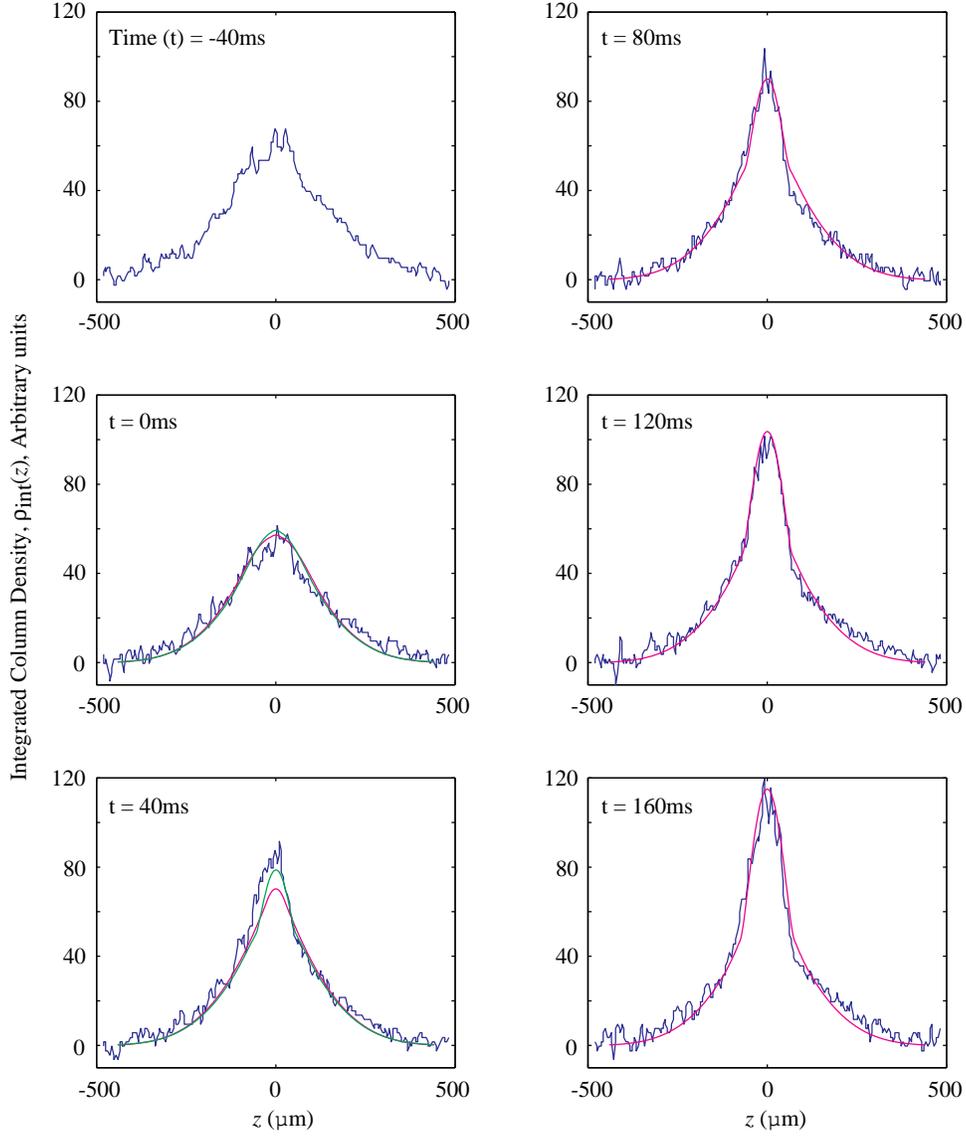,width = 13.2cm}
\end{center}
\caption{Comparison between experimental data
from~\protect\cite{MITgrowth} and theoretical spatial density distributions,
calculated using the semiclassical density distributions and the growth model
described in this paper.  The first frame shows the distribution before the
cooling `cut' below the critical temperature was performed.  The parameters
used were $T= 800{\rm nK}$ and $n_{0,f} = 9 \times 10^6$.  The (lower) red
lines show the theoretical curves using initial condition (c) described in
section~\ref{sec:initconds}.  In the second and third frames the (upper) green
curves depict the results obtained using initial condition (d) which become
essentially indistinguishable from the condition (c) lines after $40{\rm ms}$.}
\Label{fig:shapefit800nK1}
\end{figure}


\begin{figure} 
\begin{center}
\epsfig{file = 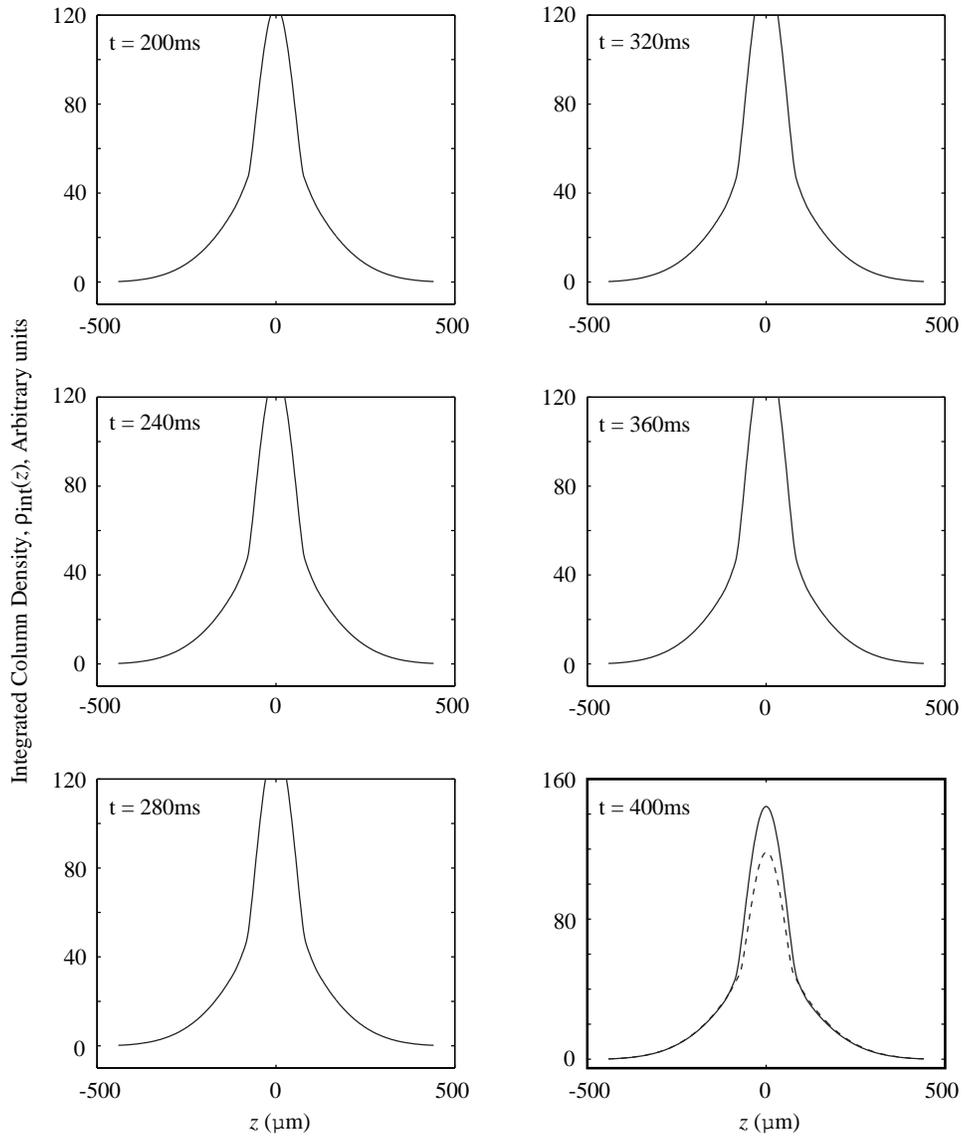,width = 13.2cm}
\end{center}
\caption{Further theoretical spatial distributions as the condensate grows. 
These graphs lead on from those in Fig.\ref{fig:shapefit800nK1} and were
obtained using the same parameters.  The scale has been chosen so as to
correspond to that in Fig.\ref{fig:shapefit800nK1}.  The last graph shows
the distribution at equilibrium ($400{\rm ms}$ - solid line) along with the
result at $160{\rm ms}$, the latest time for which the experimental data was
available.  Note the change of scale in the last graph.}
\Label{fig:shapefit800nK2}
\end{figure}
\newpage
\StartTwoColumn

\subsection{Comparison of Spatial Distributions with Experiment}

With the aid of the spatial distributions calculated using
equation~(\ref{eq:spatdist}), comparisons with the MIT growth data are
possible.  In figure 2 of~\cite{MITgrowth} the density profiles of the system
are given for a single condensate growth.  The profiles are one-dimensional
slices through the centre of the system, in the $z$ (long) axis.  The fits to
the equilibrium profile were compared in the previous section.  From this data,
MIT extracted values of $T = 800{\rm nK}$ and $n_{0,f} = 9 \times 10^6$, and
they found that the final equilibrium situation seemed to have been reached by
$160{\rm ms}$~\cite{MITpersonal}.  The theoretical spatial distribution curves
are compared to the experimental data in Fig.\ref{fig:shapefit800nK1}.  The
figure shows very good agreement with the experimental data during this $160
{\rm{ms}}$ period.  The frame at $40{\rm ms}$ shows the least good agreement,
however the agreement can be improved by the choice of different initial
conditions as is shown in Fig.\ref{fig:shapefit800nK1}.  Using initial
condition (d), described in section~\ref{sec:initconds}, the agreement with
experiment again becomes quite respectable.  It is interesting to note that the
spatial density distribution in this frame depends quite strongly on the
initial conditions, which were found to have only a very small effect on the
growth curve in section~\ref{sec:initconds}.  This is because the $40{\rm ms}$
frame is taken at a time very close to the initiation of the fast growth of the
condensate population, and the spatial distribution at this time is quite
strongly dependent on exactly when the growth does start, since any difference
in the initiation time creates a relatively large change in the occupation
numbers given that they are still quite low at this time.

The condensate growth curve which corresponds to the spatial distributions in
Fig.\ref{fig:shapefit800nK1} is shown in Fig.\ref{fig:800nKgrowth}.  The
figure shows that at $160{\rm ms}$ (the time of the last graph in
Fig.\ref{fig:shapefit800nK1}) the growth of the condensate is theoretically
only about halfway to completion.  No experimental data for this particular
growth event at times greater than $160{\rm ms}$ was published
in~\cite{MITgrowth}, although two further frames were taken at some later times
and were found by the MIT group to contain the same number of atoms as for the
$160{\rm ms}$ profile~\cite{MITpersonal}.

\begin{figure} 
\epsfig{file = 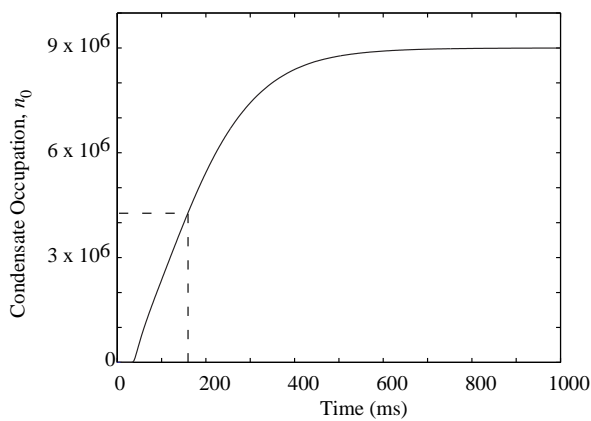,width = 8cm} 

\caption{The condensate growth curve corresponding to the situation 
in Fig.\ref{fig:shapefit800nK1}.  The dashed line indicates the 
value at $t=160{\rm ms}$ which corresponds to the final graph in 
Fig.\ref{fig:shapefit800nK1}.  Note that the growth starts to 
become significant at about the $t=40{\rm ms}$ frame.}
\Label{fig:800nKgrowth}
\end{figure}

The spatial density distributions for times later than $160{\rm ms}$ 
are shown in Fig.\ref{fig:shapefit800nK2}, for times up until 
$400{\rm ms}$ at which time equilibrium has been reached in the 
theoretical model.  The graphs in this figure show that the spatial 
distribution changes relatively little after $160{\rm ms}$ compared to 
its earlier evolution.  In particular the last graph in the figure 
compares the spatial distribution at equilibrium with that at $160{\rm 
ms}$, and it shows only a fairly small change.  The peak height has 
definitely become greater, although not by a great deal more when one 
considers the size of the fluctuations evident in the experimental 
data of Fig.\ref{fig:shapefit800nK1}.  However, more importantly, 
the spatial extent of the condensate has not changed significantly 
after $160{\rm ms}$.  The condensate number was determined in the 
majority of the MIT fits from the width of the condensate, as is given 
by equation~(\ref{condextent}).  A well known difficulty 
\cite{Wolfgang} in using this method is that, in the Thomas-Fermi 
approximation, the condensate occupation depends upon the fifth power 
of the condensate radius.  Thus, for example, a 15\%
error in the measurement of the radius will result in the value for 
$n_0$ being incorrect by a factor of $2$.  When this is considered, 
along with the fact that the experimental data will show fluctuations, 
and given the comparison between the equilibrium and $160{\rm ms}$ 
distributions, it seems perhaps possible (though in the opinion of the 
experimenters \cite{Wolfgang} very unlikely) that there was some evolution at times 
later than $160{\rm ms}$ which was not able to be measured 
experimentally to sufficient accuracy.

In conclusion therefore, more experimental data is really needed in 
order to make better comparisons, especially needed is data for times 
at which equilibrium has definitely been reached (at long $t$, 
although for very long times processes such as trap-loss may have to 
be considered).  It would also be beneficial to fit the full 
two-dimensional data, rather than the one-dimensional slices shown 
previously.  This would reduce the influence of fluctuations, and 
would lead to more weight being given to the measurements of the 
thermal cloud, which would increase the accuracy of any temperatures 
found. 

%
%

\section{Comparison with Other Theoretical Treatments}
Most of the other theoretical treatments have attempted to describe 
the formation of a condensate in a homogeneous, untrapped situation.  
No real quantitative predictions have yet emerged from any of the work 
that has been performed on trapped dilute atomic gas BEC, and so 
comparisons unfortunately will have to be qualitative at best.

\subsection{Quantum Boltzmann Equation Approach}

One of the techniques used to describe condensate growth has been the quantum
Boltzmann equation, which has been used by Snoke and
Wolfe~\cite{SW1989}, Semikoz and Tkachev~\cite{ST1995}, and Holland, Williams
and Cooper~\cite{HWC1997}, as well as forming the basis of the theory of the
kinetic stages in the work of Kagan, Svistunov and
Shlyapnikov~\cite{svist1991,KSS1992}.  Although the theory described in this
paper was developed from the Quantum Kinetic theory~\cite{QK1,QK2,QK3}, it
turns out that essentially the same equations may be obtained by modifying
the quantum Boltzmann equation approach as follows:
\Item[(i)] The quantum Boltzmann equation in an ergodic form is used, a form
similar to that used in~\cite{HWC1997}
\begin{eqnarray}
\frac{\partial f(e_n)}{\partial t} 
&= &
{8ma^2\omega^2 \over \pi\hbar}\sum_{e_m,e_p,e_q}
\delta(\Delta E) g(e_{\rm min}) \times\nonumber \\
&& \Big[f(e_p)f(e_q) (1+f(e_m)) (1+f(e_n)) \nonumber \\
&&  \quad - f(e_m)f(e_n) (1+f(e_p)) (1+f(e_q))\Big],
\nonumber\\
\end{eqnarray}
where $n_k = g_kf(e_k)$ is the number of particles with energy $e_k$,
$e_{{\min}} =\min(e_{m},e_{n},e_{p},e_{q})$ and $\Delta E = e_m +e_n-e_p-e_q$. 

\Item[(ii)] The energy levels in the condensate band are modified as discussed
in section~\ref{sec:modenergies}, in order to account for the mean field
interactions due to the presence of the condensate.

\Item[(iii)] The levels in the non-condensate band are summed over and assumed
to be time independent.  This allows much larger, and realistic sized, systems
to be modelled as opposed to the $100$-$1000$ atom systems typically simulated
in previous attempts.

\Item[(iv)] Collisions between two particles which were both initially in the
condensate band were neglected, this is a valid approximation if the vast
majority of particles are found in the non-condensate band. 
\Item[]%
Using these modifications, and the rates for the scattering and growth
processes found from quantum kinetic theory (see sec.\ref{ScattProc}), the quantum Boltzmann equation will
give rise to the set of differential equations~(\ref{totalgrowth}) whose
solutions provided the results in this paper.

Of the above references, only the work of Holland, Williams and Cooper
conducted any simulations for the growth of a trapped condensate.  They found
that their simulations of condensate occupation number evolution behaved as
$n_0 = n_{0,f}(1-e^{-t/\tau})$, where $\tau$ was a fitted parameter.  A
function of this form can be made to fit the results obtained by our model
reasonably well, provided that an initiation time is allowed for, as was
anticipated might be necessary in~\cite{HWC1997}.  The same functional form was
obtained using the quantum Boltzmann master equation approach in~\cite{QK2} by
Jaksch \emph{et al.}

It should be noted that, although the quantum kinetic description for the
growth of the \emph{mean} occupation numbers turns out to also be described by
a modified quantum Boltzmann equation approach, the full quantum kinetic theory
treats aspects of condensate dynamics which are not accessible via the quantum
Boltzmann equation. Such aspects include: the treatment of fluctuations, phase
and phase decoherence, and the inclusion of Bogoliubov-like quasiparticle
states. 

\subsection{Comparison with Work of Kagan \emph{et al.}}

Major theoretical work into the growth of condensate in recent years 
has been performed by Kagan, Svistunov, and Shlyapnikov.  They divided 
the growth into three stages, the first of which was a kinetic stage 
described by Svistunov in~\cite{svist1991}.  Svistunov predicted the 
formation of a particle-flux wave in energy space during the initial 
stages, which transports particles towards lower energy states.  The 
arrival of this wave at the lowest energy state at the \emph{critical 
time} would give rise to an energy distribution function of the form 
$f(E) \propto E^{-7/6}$.  After the critical time this behaviour would 
be lost due to a particle-flux wave propagating to higher energies.  
The simulations of Semikoz and Tkachev~\cite{ST1995} showed this 
behaviour to some extent, although they found that the behaviour at 
the critical time was $f(E) \propto E^{-1.24}$.

The work by Svistunov in~\cite{svist1991} related to the case of homogeneous
systems, however he recently reworked his methodology to consider a gas
confined by a harmonic trapping potential.  In this case he
found~\cite{Svistpersonal} that the dependence at the critical time now tended
towards $f(E) \propto E^{-5/3}$.

\begin{figure} 
\epsfig{file = 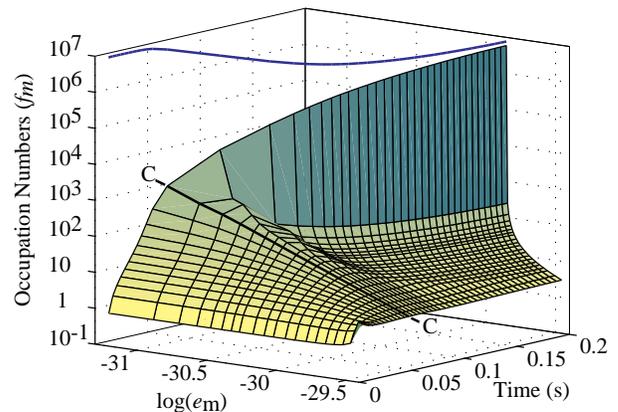,width = 8cm}

\caption{The growth of a condensate of 9 million atoms as predicted 
by the model in this paper.  Plotted is $f(e_m)$, the population per 
individual level with energy $e_m$, and the energies of the levels as 
functions of time.  The values of $f(e_m)$ and $e_m$ are plotted on 
log scales.  The lines almost parallel to the time axis are not lines 
of constant energy, but rather lines of constant level number, whose 
energies change with time.  Not all levels are shown in order to make 
the behaviour legible.  The solid black curve represents the energy 
of the condensate level $\log_{10}[\mu_N(n_0)]$ as a function of 
time.  The bold line labelled C--C represents the critical time, at 
which the energy distribution has the form $f(E) \propto E^{-1.61}$.  
The temperature of the system was 
$800{\rm nK}$.}
\Label{fig:EFTmesh}
\end{figure}
In Fig.\ref{fig:EFTmesh} the results our model of condensate growth are
shown in a somewhat different form.  The occupation numbers for both the
condensate level and excited states are plotted as a function of their energy
and the time (note the logarithmic scales). From this graph several points can
be noted, firstly the front corner shows quite how rapidly the initial
conditions, however arbitrary, are smoothed out by scattering processes, and
the discontinuous initial conditions rapidly approach a realistic
distribution.  The growth of the condensate is rather small up to the point
labelled as the critical time, after which the condensate grows enormously. 
The populations of the excited states approach equilibrium very rapidly after
the critical time, much more rapidly than the condensate level does.

As the critical time is approached, the distribution approaches a straight line
as shown in figure
~\ref{fig:EFTmesh}
.  At the critical
time, when this distribution is linear with the logarithm of the energy, the
energy dependence was found to be of the form $f(E) \propto E^{-1.61}$, which
is in good agreement with the prediction of Svistunov of $E^{-5/3}$.

As far as estimates of timescales are concerned, Kagan, Svistunov and
Shlyapnikov predicted that the evolution of the system up until initiation
would occur on the timescale of the classical collision time $\tau_0$, that the
initiation of the condensate growth would occur on a much faster timescale, and
that the final growth would occur on the timescale necessary for the
annihilation of vortices in homogeneous gases, and the decay of fluctuations in
the phase of the condensate.

In the treatment of Kagan, Svistunov and Shlyapnikov'the timescale 
for the first kinetic stage was postulated for a homogeneous gas, 
however a comparison can still be made.  A first estimate of $\tau_0 
= (\sigma\bar{n}v_T)^{-1}$ can be obtained by using the classical 
value for $v_T$, the mean thermal velocity in a gas, of $v_T = 
\sqrt{2kT/m}$, and the cross section defined by $\sigma = 8\pi a^2$.  
The value for $\bar{n}$, the mean density, will be taken as the 
density of the non-condensate particles in the centre of the trap 
given by
\begin{eqnarray}
\bar{n} &=& \left[{mkT\over 2\pi\hbar^2}\right]^{3/2}G_{3/2}
\left({V_T(0)\over kT}\right)
\end{eqnarray}
 (see equation~\ref{eq:g32}) where $G_\alpha(z) = \sum_{q=1}^\infty
e^{-qz}/q^\alpha$.  Taking $V_T(0) = 0$, and using a
temperature of $900{\rm nK}$, the collision time is $\tau_0 = 27{\rm ms}$. 
This temperature corresponds to that of the growth in
Fig.\ref{fig:sampletotal}, which shows that the time until initiation is of
the order of $2\tau_0$. Thus our treatment does agree with the picture of Kagan
\emph{et al.} in that this stage occurs
over the order of a few $\tau_0$.

A note about the collision times is needed here.  In examining the validity of
the ergodic approximation Jaksch \emph{et al}.\ found that it was valid only
for quantities averaged over about 10 collision times~\cite{QK2}.  The above
timescale $\tau_0$ is \emph{not} the timescale over which collisions occur in
the condensate system in reality.  It is rather the classical collision time
for a classical gas in equilibrium below the critical temperature but with
\emph{no} condensate present, and is obviously artificial.  Once the condensate
begins to form the density increases significantly, and the actual mean
collision time was found in~\cite{QK2} to be more than two orders of magnitude
smaller than $\tau_0$.  Thus the ergodic assumption should still be valid for
our treatment, even given the large value of $\tau_0$.

The timescales found by Kagan \emph{et al.} for the second and third stages of
evolution have only been determined for the case of a homogeneous gas, and so
accurate comparisons with our model for these stages are not able to be
performed.  Our model does agree that the initiation stage occurs on a much
faster timescale than the first kinetic stage.  The presence of
vortices has not been considered in our treatment, nor has any consideration
been given to the phase fluctuations, and so comparisons cannot be made with the
third stage of evolution in the description of Kagan \emph{et al.}

%

%
%

\section{Conclusions}


In this paper a model of the growth of a condensate, derived from quantum
kinetic theory~\cite{QK1,QK2,QK3}, is presented.  The main improvements over
the simpler description in~\cite{BosGro} are the more accurate calculation of
the growth rate factor $W^+(n_0)$, the consideration of the time dependence of
the lower energy levels, and the inclusion of the scattering of particles
between these levels.  The modified $W^+$ factors have the greatest effect,
generally increasing the rate of growth by a factor of about 3, dependent on
the exact parameters.  The inclusion of the other levels and their scattering
also leads to an increase in the rate of growth, mainly by reducing the amount
of time taken for the initiation of the growth, that is the time before the
stimulated growth processes, due to the Bose statistics of the system, become
dominant.

The model describes the evolution of time-dependent energy levels in the lowest
states, coupled with a time independent thermal bath of atoms occupying the
higher energy levels. 

The results give growth curves whose shapes are approximately those given
by
\begin{eqnarray}
n_0 = \left\{
\begin{array}{ll}
0 & \mbox{for} \ t < t_i \\
n_{0,f}(1-e^{-(t-t_i)/\tau}) & \mbox{for} \ t>t_i
\end{array} \right.
\end{eqnarray}
where $t_i$ is some initiation time.  This form agrees with the general form of
the results of Holland \emph{et al.}~\cite{HWC1997} and Jaksch \emph{et
al.}~\cite{QK2} once an initiation time is allowed.  The results also seem to
be in qualitative agreement with features of the description proposed by Kagan,
Svistunov and Shylapnikov.

The results are not very sensitive to the
exact nature in which the mean-field effects on the lower levels were accounted
for, so long as the energies of the very lowest levels were altered in a
consistent fashion.  The initial conditions used did not have a large effect on
the growth curves, however they can be important when the spatial density
profiles of the system are calculated for comparison with experiment.  

The evolution of the model depends upon approximations made for the rate
factors $\Gamma(T)$ and $W^{++}_m(n_0)$.  The results obtained show that the
growth is not very different as long as the actual value of $\Gamma(T)$ falls
within about a factor of 10 of the approximation, and as long as the actual
values of $W^{++}_m(n_0)$ lies within a factor of 2 of $W^+(n_0)$.


Overall the rates of growth predicted now agree very well with the 
growth rates measured from experimental data \cite{MITgrowth}.  
However at lower temperatures the trend in growth rates shows some 
divergence, with the experimental rates becoming quicker whilst the 
predicted rates become slower.  This may be a result of the 
substantial cooling necessary to achieve these temperatures, giving 
rise to a highly non-equilibrium system which is inadequately 
described by our model.

The extraction from experimental data of energy level populations is 
fraught with difficulties, leading to large uncertainties.  It was 
found to be better to compare theoretical predictions of the 
evolution of the spatial density distribution of the system as the 
condensate to those experimentally measured, and the results in 
sect.\ref{spatial} appear to give good agreement with the spatial 
shapes given in the MIT data.  They also show that, once a large 
condensate occupation has been reached, if the condensate occupation 
changes by a significant amount the corresponding change in the 
spatial density distribution of the system is relatively small.

With the aid of more raw experimental data to fit, a determination of 
exactly how accurate the predictions of this model are, conclusions 
as to the exact initial conditions present, and more accurate 
approximations of $\Gamma(T)$ and $W^{++}_m(n_0)$ may be able to be 
reached.  


Further work which could be undertaken within the framework of this 
model includes:

\Item[i)] An accurate determination of the rate factors $\Gamma(T)$ 
and $W^{++}_m(n_0)$ analytically, or at least finding constraints on 
their values by comparison with more experimental data.

\Item[ii)] An inclusion of some Bogoliubov phonon-like quasiparticle 
nature in the description of the lower energy levels, since all 
excited levels in this paper were treated as Hartree-Fock 
particle-like quaiparticles, which will be valid for most of the 
higher levels but not for the lower energy excitations.

\Item[iii)] A consideration of the fluctuations in occupation 
numbers.  These may be significant in determining the initiation 
time, which is when the occupation of the condensate level becomes 
large enough for the stimulated growth processes to take over.

\Item[iv)] As the model stands, the non-condensate band `bath' of 
atoms is treated as being time independent.  A major extension of the 
model would be to include the dynamics of the non-condensate band in 
the evolution.  Extension to include a time-dependent bath will be 
treated elsewhere \cite{QKVII}.

\acknowledgments
We would like to thank especially Wolfgang Ketterle and Hans-Joachim 
Miesner for discussions regarding the interpretation of their data, 
as well as Yuri Kagan and Boris Svistunov for discussions concerning their 
work on the kinetic theory of condensate initiation.  The research 
was supported by the Royal Society of New Zealand unde Marsden fund 
Contracts PVT-603 and PVT-902.

\EndTwoColumn
\end{document}